\documentclass[runningheads]{llncs}
\usepackage{color}
\usepackage{amsmath}
\usepackage{amssymb} 
\usepackage{wasysym}
\usepackage{graphicx}
\usepackage{xspace}
\usepackage{subfig}
\usepackage[export]{adjustbox}
\usepackage{stmaryrd}
\usepackage{enumitem}
\usepackage{tikz}
\usepackage{tikz-layers}
\usetikzlibrary{patterns}
\usetikzlibrary{decorations.markings}
\usepackage{xspace}
\usepackage{pgfplotstable}
\usepackage{booktabs}
\usepackage{latexsym}
\usepackage{listings}

\usepackage[T1]{fontenc}
\usetikzlibrary{automata, positioning, arrows}

\usetikzlibrary{decorations.pathmorphing}
\usetikzlibrary{calc}

\makeatletter
\DeclareRobustCommand{\cev}[1]{%
  \mathpalette\do@cev{#1}%
}
\newcommand{\do@cev}[2]{%
  \fix@cev{#1}{+}%
  \reflectbox{$\m@th#1\vec{\reflectbox{$\fix@cev{#1}{-}\m@th#1#2\fix@cev{#1}{+}$}}$}%
  \fix@cev{#1}{-}%
}
\newcommand{\fix@cev}[2]{%
  \ifx#1\displaystyle
    \mkern#23mu
  \else
    \ifx#1\textstyle
      \mkern#23mu
    \else
      \ifx#1\scriptstyle
        \mkern#22mu
      \else
        \mkern#22mu
      \fi
    \fi
  \fi
}

\makeatother

\newcommand{\msep}{\,\mid\,}

\newcommand{\slcs}{{\tt SLCS}}
\newcommand{\slcsG}{{\tt SLCS}$_{\gamma}$}
\newcommand{\slcsE}{{\tt SLCS}$_{\eta}$}

\newcommand{\voxlogica}{{\tt VoxLogicA}}
\newcommand{\polylogica}{{\tt PolyLogicA}}
\newcommand{\imgql}{{\tt ImgQL}}
\newcommand{\polyvisualizer}{{\tt PolyVisualizer}}

\newcommand{\SET}[1]{\{#1\}}
\newcommand{\ZET}[2]{\SET{#1 \,|\, #2}}
\newcommand{\pws}[1]{\mathbf{2}^{#1}}
\newcommand{\nats}{\mathbb{N}}
\newcommand{\reals}{\mathbb{R}}

\newcommand{\cnv}[1]{#1^{-}}
\newcommand{\dircnv}[1]{#1^{\pm}}

\newcommand{\plm}{$\pm$}

\newcommand{\relint}[1]{\widetilde{#1}}
\newcommand{\ap}{{\tt PL}}

\newcommand{\map}{\mathbb{F}}
\newcommand{\peval}{\calV}

\newcommand{\sibis}{\sim_{\resizebox{0.2cm}{!}{$\triangle$}}}
\newcommand{\wsibis}{\approx_{\resizebox{0.2cm}{!}{$\triangle$}}}

\newcommand{\plmbis}{\sim_{\pm}}
\newcommand{\wplmbis}{\approx_{\pm}}

\newcommand{\form}{\Phi}
\newcommand{\ltrue}{{\tt true}}

\newcommand{\lneg}{\neg}

\newcommand{\slcsGeq}{\equiv_{\gamma}}
\newcommand{\slcsEeq}{\equiv_{\eta}}

\newcommand{\bDiamond}{\mathbin{\Diamond}}

\makeatletter
\newcommand\bigDiamond{\mathop{\mathpalette\bigDi@mond\relax}}
\newcommand\bigDi@mond[2]{%
  \vcenter{\hbox{\m@th
    \scalebox{\ifx#1\displaystyle 2\else1.2\fi}{$#1\Diamond$}%
  }}%
}

\makeatother

\newcommand{\dirnear}{\stackrel{\rightarrow}{\bDiamond}}
\newcommand{\cnvnear}{\stackrel{\leftarrow}{\bDiamond}}

\newcommand{\closure}{\calC}

\newcommand{\calC}{\mathcal{C}}

\newcommand{\calF}{\mathcal{F}}

\newcommand{\calP}{\mathcal{P}}

\newcommand{\calV}{\mathcal{V}}

\newcommand{\CYAN}[1]{\textcolor{cyan}{#1}}
\newcommand{\RED}[1]{\textcolor{red}{#1}}

\newcounter{dgnot} % Notes from Diego
\newenvironment{dgnot}[1][]{\refstepcounter{dgnot}\par\medskip
   \noindent \textbf{\RED{NfDiego~\thedgnot.}  #1} \rmfamily}{\medskip}

\newcounter{mknot} % Notes from Mieke
\newenvironment{mknot}[1][]{\refstepcounter{mknot}\par\medskip
   \noindent \textbf{\CYAN{NfMieke~\themknot.}  #1} \rmfamily}{\medskip}

\begin{document}
\title{Practical Exploration of Polyhedral\\ Model Checking\thanks{The authors are listed in alphabetical order, as they equally contributed to the work presented in this paper.}}
% AUTHOR INFO FOR FINAL VERSION
\author{
Yuri Andriaccio\inst{1} \and
Vincenzo Ciancia\inst{2} \and %\orcidID{0000-0003-1314-0574} \and
Diego Latella\inst{3} \and %\orcidID{0000-0002-3257-9059} \and
Mieke Massink\inst{2} %\orcidID{0000-0001-5089-002X}
}

\authorrunning{}
% INSTITUTE INFO FOR FINAL VERSION
\institute{
University of Pisa, Pisa, Italy\\
\email{y.andriaccio@studenti.unipi.it}
\and
Istituto di Scienza e Tecnologie dell'Informazione ``A. Faedo'', Consiglio Nazionale delle Ricerche, Pisa, Italy\\
\email{\{Vincenzo.Ciancia, Mieke.Massink\}@cnr.it}\\
\and
Formerly with Istituto di Scienza e Tecnologie dell'Informazione ``A. Faedo'', Consiglio Nazionale delle Ricerche, Pisa, Italy\\
\email{diego.latella@actiones.eu}\\
}

\maketitle              % typeset the header of the contribution
\begin{abstract}
This work explores the potential of spatial model checking of polyhedral models on a number of selected examples. In computer graphics polyhedral models can be found in the form of triangular surface meshes of tetrahedral volume meshes which are abundant. Spatial model checking is used to analyse spatial properties of interest of such models expressed in a suitable spatial logic. The original contributions of this paper are twofold. First we illustrate how a polyhedral model can be enriched by adding the outcome of one model checking session as an atomic proposition to the original model. This is useful as it provides a way to reduce the length of formulas to check on such models and to obtain more insightful results when these models are used for graphical visualisation. Second we show that this form of enrichment also enables practical model minimisation providing deeper insights in the basic spatial structure of the model in terms of the spatial logic properties it enjoys. This work is performed in the context of the geometric spatial model checker \polylogica{}, the visualizer \polyvisualizer\ and the polyhedral semantics of the Spatial Logic for Closure Spaces \slcs{.}
\end{abstract}

\section{Introduction}

\begin{figure}[h!]
\centering
\includegraphics[width=0.5\textwidth]{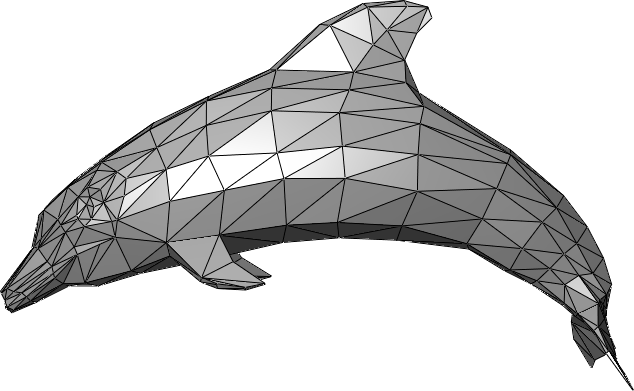}
\caption{A triangular surface mesh of a dolphin~\cite{Wik23}.}\label{fig:dolphin}
\end{figure}

An interesting class of spatial models are polyhedra. These form the mathematical basis for the visualisation of objects in {\em continuous} space. In the real world we find such structures in domains that exploit mesh-processing such as in computer graphics. Fig.~\ref{fig:dolphin} shows a simple example of a surface triangular mesh of a dolphin. Such triangular surface meshes and tetrahedral volume meshes can be very large. Spatial model checking for such structures can be used to highlight or identify interesting aspects of the structures.

In~\cite{Be+22} we addressed the theory and model checking algoritms for such continuous spaces and developed the \polylogica\ model checker for polyhedral model checking and an associated visualiser \polyvisualizer\ for the graphical presentation of the model checking results. More specifically, we considered {\em polyhedral models} and \slcsG, a variant of \slcs{} for such models. A polyhedral model is composed of a {\em polyhedron} and a {\em valuation function}. 
A polyhedron $|K|$ is the set union of all the  components of a simplicial complex $K$, which, in turn, is a collection of {\em simplexes} in  $\reals^n$, for some dimension $n$, satisfying certain construction constraints. 
For our purposes, we set $n$ to $3$, so that, by definition, the kind of simplexes we are concerned with
are just points, segments, triangles and tetrahedra. 

The original contributions of this paper are twofold. As a first contribution, we illustrate how a polyhedral model can be enriched by adding the outcome of one model checking session as an atomic proposition to the original model. This is useful as it provides a way to reduce the length of formulas to check on such models and to obtain more insightful results when these models are used for graphical visualisation. As a second contribution, we show that this form of enrichment also enables practical model minimisation providing deeper insights in the basic spatial structure of the model in terms of the spatial logic properties it enjoys. This work is performed in the context of the geometric spatial model checker \polylogica{}, the visualizer \polyvisualizer\ and the polyhedral semantics of the Spatial Logic for Closure Spaces \slcs{.} We first briefly recall an earlier example from~\cite{Be+22} to illustrate polyhedral model checking on a synthetic example so that the reader gets acquainted with this novel form of spatial model checking. We then illustrate the contributions using two different case studies originating from the large set of available mesh models in Wavefront .obj format in the field of computer graphics. 

\subsubsection{Related work.} Model checking of simplicial complexes is also addressed in~\cite{LoQ23}. Therein a logic is defined that shares a similar syntax as \slcs\ but is provided with a different semantics than the one used in the present paper. The focus of their work is on modelling and analysis of groups of interactions and their higher order relationships. In particular, in their work the domain upon which formulas are interpreted are (sets of) simplexes and not (sets of) points in a polyhedra as in the present paper. 

From the tools point of view, the python library pymeshlab~\cite{pymeshlab}, which is able to programmatically modify 3D meshes based on pre-built operators (mostly traditional 3D imaging filters) is somewhat related to our work. However, a direct comparison of this library and polyhedra model checking would be misleading as the latter uses a declarative language based on \slcs{,} with automatic parallelisation, and automatic memoization (caching) of intermediate results instead of providing a library of pre-built operators. 

\subsubsection{Synopsis.} Section~\ref{sec:BackAndNotat} recalls notation and provides background on relevant concepts concerning polyhedral models, spatial logic, model checking and minimisation. Section~\ref{sec:examples} presents a selection of examples of surface and volume meshes and illustrates various ways in which spatial polyhedral model checking can be used on such models. Section~\ref{sec:conclusion} concludes the paper.

\section{Background and Notation}\label{sec:BackAndNotat}

We briefly recall notation and relevant background information on the language \slcsG, its polyhedral and poset models, and
the truth-preserving map $\map$ between these models.

\subsubsection{General notation.} For sets $X$ and~$Y$, a function $f:X \to Y$, and subsets
$A \subseteq X$ and $B \subseteq Y$ we define $f(A)$ and~$f^{-1}(B)$
as $\ZET{f(a)}{a \in A}$ and $\ZET{a}{f(a) \in B}$, respectively.
The  \emph{restriction} of~$f$ on~$A$ is denoted by~$f|A$.
The powerset of~$X$ is denoted by~$\pws{X}$.
For a binary relation $R \subseteq {X \times X}$ we let
$\cnv{R} = \ZET{(y,x)}{(x,y)\in R}$ denote its converse and let
$\dircnv{R}$ denote $R \, \cup \cnv{R}$. For partial orders~$\preceq$
we will use the standard notation~$\succeq$ for~$\cnv{\preceq}$ and
$x \prec y$ whenever $x \preceq y$ and $x \neq y$ (and similarly
for~$x \succ y$). If $R$ is an equivalence relation on $A$, we let
$A{/R}$ denote the {\em quotient} of $A$ via $R$.
In the remainder of the paper we assume that a set~$\ap$ of
\emph{proposition letters} is fixed. The sets of natural numbers and
of real numbers are denoted by $\nats$ and~$\reals$, respectively. We
use the standard interval notation: for $x,y \in \reals$ we let
$[x,y]$ be the set $\ZET{r \in \reals}{x \leq r \leq y}$,
$[x,y) = \ZET{r\in \reals}{x \leq r < y}$, and so on. Intervals
of~$\reals$ are equipped with the Euclidean topology inherited
from~$\reals$. We use a similar notation for intervals over~$\nats$:
for $n,m \in \nats$, $[m;n]$ denotes the set
$\ZET{i \in \nats}{m \leq i \leq n}$,
$[m;n) = \ZET{i \in \nats}{m \leq i < n}$, and so on.

Below we recall some basic notions, assuming that the
reader is familiar with topological spaces, Kripke models, and
posets.

\subsubsection{Polyhedral Models and Cell Poset Models.}
A \emph{simplex} $\sigma$ of dimension $d$ is the convex hull of a set 
$\SET{\mathbf{v_0},\ldots, \mathbf{v_d}}$ of
$d+1$~affinely independent points in~$\reals^m$, with $d \leq m$,
i.e.\
$\sigma = \ZET{ \lambda_0\mathbf{v_0} + \ldots +
  \lambda_d\mathbf{v_d}}{\lambda_0,\ldots,\lambda_d \in [0,1]\mbox{
    and } \sum_{i=0}^{d} \lambda_i = 1}$. For instance, a segment~$AB$
together with its end-points $A$ and~$B$
is a simplex in~$\reals^m$, for$~m \geq 1$.  Any subset of the set $\SET{\mathbf{v_0},\ldots, \mathbf{v_d}}$ of
points characterising a simplex~$\sigma$ induces a simplex~$\sigma'$ in turn,
and we write $\sigma' \sqsubseteq \sigma$, noting that
$\sqsubseteq$~is a partial order, e.g.\ 
$A \sqsubseteq A \sqsubseteq AB$, $B \sqsubseteq B \sqsubseteq AB$ and $AB \sqsubseteq AB$. 

The \emph{relative interior} $\relint{\sigma}$ of a simplex~$\sigma$ is the
same as $\sigma$ ``without its borders'', i.e.\ the set
$\ZET{ \lambda_0\mathbf{v_0} + \ldots +
  \lambda_d\mathbf{v_d}}{\lambda_0,\ldots,\lambda_d \in (0,1]\mbox{
    and } \sum_{i=0}^{d} \lambda_i = 1}$. For instance, the open
segment~$\relint{AB}$, without the end-points $A$ and~$B$ is the
relative interior of segment~$AB$. The relative interior of a simplex
is often called a~\emph{cell} and is equal to the topological interior
taken inside the affine hull of the simplex.\footnote{But note that
  the relative interior of a simplex composed of just a single point
  is the point itself and not the empty set.}  The partial
order~$\sqsubseteq$ is reflected on cells:
$\relint{\sigma_1} \preccurlyeq \relint{\sigma_2}$ if and only if
$\sigma_1 \sqsubseteq \sigma_2$. 
Obviously, $\preccurlyeq$ is a partial order. In the above
example, we have
$\relint{A}\preccurlyeq \relint{A} \preccurlyeq \relint{AB}, \relint{B}\preccurlyeq \relint{B} \preccurlyeq \relint{AB},$ and
$\relint{AB}\preccurlyeq \relint{AB}$.

A \emph{simplicial complex}~$K$ is a finite collection of simplexes
of~$\reals^m$ such that: (i) if $\sigma \in K$ and
$\sigma' \sqsubseteq \sigma$ then also $\sigma' \in K$; (ii) if
$\sigma, \sigma' \in K$ and $\sigma \cap \sigma' \not=\emptyset$, then
$\sigma \cap \sigma' \sqsubseteq \sigma$ and
$\sigma \cap \sigma' \sqsubseteq \sigma'$.
The \emph{cell poset} of simplicial complex~$K$ is
$(\relint{K},\preccurlyeq)$ where $\relint{K}$ is the set
$\ZET{\, \relint{\sigma}}{\sigma \in K}$, and $\preccurlyeq$ is the union of the partial orders on the cells of the simplexes of $K$. 
It is easy to see that $\preccurlyeq$ itself is a partial order, due to the geometrical nature of cells.

The {\em polyhedron}~$|K|$ of~$K$
is the set-theoretic union of the simplexes in~$K$. Note that
$|K|$~inherits the topology of~$\reals^m$ and that $\relint{K}$ forms a partition of polyhedron $|K|$.
Note furthermore that different simplicial complexes can give rise to the same polyhedron.

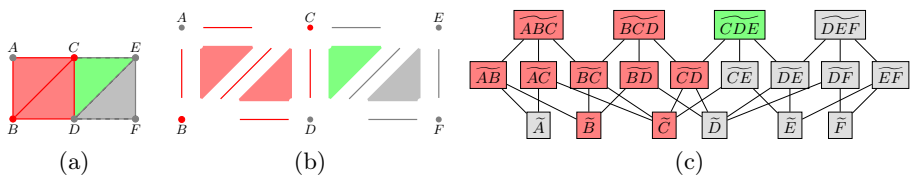
\begin{figure}[h]
\subfloat[]{\label{subfig:PolyhedronNoPathCompressed}
\resizebox{0.8in}{!}
{
\begin{tikzpicture}[scale=1.4,label distance=-2pt]
	    \tikzstyle{point}=[circle,draw=black,fill=white,inner sep=0pt,minimum width=4pt,minimum height=4pt]
	    \node (p0)[point,draw=red,label={270:$B$}] at (0,0) {};
	    	\filldraw [red] (p0) circle (1.25pt);
%	    \node (p1)[point,draw=red,label={ 90:$A$}] at (0,1) {};
	    \node (p1)[point,draw=gray,label={ 90:$A$}] at (0,1) {};
%	    	\filldraw [red] (p1) circle (1.25pt);
	    	\filldraw [gray] (p1) circle (1.25pt);
	    \node (p2)[point,draw=gray,label={270:$D$}] at (1,0) {};
	    \node (p3)[point,draw=red,label={ 90:$C$}] at (1,1) {};
	    	\filldraw [red] (p3) circle (1.25pt);
	    \node (p4)[point,draw=gray,label={270:$F$}] at (2,0) {};
	    \node (p5)[point,draw=gray,label={ 90:$E$}] at (2,1) {};

	    \draw [red   ,thick](p0) -- (p1);
	    \draw [red   ,thick](p0) -- (p2);
	    \draw [red   ,thick](p0) -- (p3);
	    \draw [red   ,thick](p1) -- (p3);
	    \draw [red   ,thick](p2) -- (p3);	    
    \draw [dashed      ](p2) -- (p4);
    \draw [dashed      ](p2) -- (p5);
    \draw [dashed      ](p3) -- (p5);
    \draw [dashed      ](p4) -- (p5);
    \draw [gray,thick](p2) -- (p4);
    \draw [gray,thick](p4) -- (p5);
    \draw [gray,thick](p2) -- (p5);
    \draw [gray,thick](p3) -- (p5);
	        
	    \begin{scope}[on background layer]
	    \fill [fill=red!50  ](p0.center) -- (p1.center) -- (p3.center);
	    \fill [fill=red!50  ](p0.center) -- (p3.center) -- (p2.center);
	    \fill [fill=green!50](p2.center) -- (p3.center) -- (p5.center);
            \fill [fill=gray!50](p2.center) -- (p4.center) -- (p5.center);    	    
            \end{scope}

    \filldraw [gray] (p2) circle (1.25pt);
    \filldraw [gray] (p4) circle (1.25pt);
    \filldraw [gray] (p5) circle (1.25pt); 
	\end{tikzpicture}
	}
}
\subfloat[]{\label{subfig:PolyhedronNoPathCellsCompressed}
\resizebox{1.5in}{!}
{
\begin{tikzpicture}[scale=1.3,label distance=-2pt]
	    \tikzstyle{point}=[circle,fill=white,inner sep=0pt,minimum width=4pt,minimum height=4pt]
	    \node (p0S0d)[point,draw=red,fill=red,label={270:$B$}] at (0,0) {};
	    \node (p0S1d)[point] at (0.33,0) {};
	    \node (p0S2d)[point] at (0.66,0) {};
	    \node (p0S3d)[point] at (0.99,0) {};
	    \node (p0S0u)[point] at (0,0.33) {};
	    \node (p0S1u)[point] at (0.33,0.33) {};
	    \node (p0S2u)[point] at (0.66,0.33) {};
	    \node (p0S3u)[point] at (0.99,0.33) {};
	    
	    \node (p1S0d)[point] at (0,1.33) {};
	    \node (p1S1d)[point] at (0.33,1.33) {};
	    \node (p1S0u)[point,fill=gray,label={90:$A$}] at (0,1.66) {};
	    \node (p1S1u)[point] at (0.33,1.66) {};
	    
	    \node (p2S0u)[point] at (1.99,0.33) {};
	    \node (p2S0d)[point] at (1.99,0) {};
	    \node (p2S1u)[point] at (2.32,0.33) {};
	    \node (p2S1d)[point,fill=gray,label={270:$D$}] at (2.32,0.0) {};
	    \node (p2S2u)[point] at (2.65,0.33) {};
	    \node (p2S3u)[point] at (2.98,0.33) {};
	    \node (p2S4u)[point] at (3.31,0.33) {};
	    \node (p2S4d)[point] at (3.31,0.0) {};
	    
	    \node (p3S0u)[point] at (1.33,1.66) {};
	    \node (p3S1u)[point,fill=red,label={90:$C$}] at (2.32,1.66) {};
	    \node (p3S0d)[point] at (1.33,1.33) {};
	    \node (p3S1d)[point] at (1.66,1.33) {};
	    \node (p3S2u)[point] at (2.65,1.66) {};
	    \node (p3S3u)[point] at (2.98,1.66) {};
	    \node (p3S2d)[point] at (1.99,1.33) {};
	    \node (p3S3d)[point] at (2.32,1.33) {};
	    \node (p3S4d)[point] at (2.65,1.33) {};
	    
	    \node (p4S0u)[point] at (4.31,0.33) {};
	    \node (p4S0d)[point] at (4.31,0.0) {};
	    \node (p4S1u)[point] at (4.64,0.33) {};
	    \node (p4S1d)[point,fill=gray,label={270:$F$}] at (4.64,0.0) {};	    
	    
	    \node (p5S0u)[point] at (3.65,1.66) {};
	    \node (p5S1u)[point,fill=gray,label={90:$E$}] at (4.64,1.66) {};
	    \node (p5S0d)[point] at (3.65,1.33) {};
	    \node (p5S1d)[point] at (3.98,1.33) {};
	    \node (p5S2d)[point] at (4.31,1.33) {};
	    \node (p5S3d)[point] at (4.64,1.33) {};
	    
	    \draw [red,thick](p0S0u) -- (p1S0d);
	    \draw [red,thick](p1S1u) -- (p3S0u);
	    \draw [red,thick](p0S2u) -- (p3S1d);
	    \draw [red,thick](p0S3d) -- (p2S0d);
	    \draw [red,thick](p3S3d) -- (p2S1u);
	    \draw [gray,thick](p2S3u) -- (p5S1d);
	    \draw [gray,thick](p3S2u) -- (p5S0u);
	    \draw [gray,thick](p2S4d) -- (p4S0d);
	    \draw [gray,thick](p5S3d) -- (p4S1u);
	    	    
	    \begin{scope}[on background layer]
	    \fill [fill=red!50  ](p0S1u.center) -- (p1S1d.center) -- (p3S0d.center);
 	    \fill [fill=red!50  ](p0S3u.center) -- (p3S2d.center) -- (p2S0u.center);
	    \fill [fill=green!50  ](p2S2u.center) -- (p3S4d.center) -- (p5S0d.center);
	    \fill [fill=gray!50  ](p2S4u.center) -- (p5S2d.center) -- (p4S0u.center);    
            \end{scope}
	\end{tikzpicture}
	}
}
\subfloat[]{\label{subfig:PolyhedronNoPathPosetCompressed}
\resizebox{2.3in}{!}
{
\begin{tikzpicture}[scale=20, every node/.style={transform shape}]
    \tikzstyle{kstate}=[rectangle,draw=black,fill=white]
    \tikzset{->-/.style={decoration={
		markings,
		mark=at position #1 with {\arrow{>}}},postaction={decorate}}}
    
    \node[kstate,fill=red!50  ] (P0) at (  1,0) {$\relint{B}$};
    \node[kstate,fill=lightgray!50  ] (P1) at (  0,0) {$\relint{A}$};
    \node[kstate,fill=lightgray!50] (P2) at (3.5,0) {$\relint{D}$};
    \node[kstate,fill=red!50  ] (P3) at (2.5,0) {$\relint{C}$};
    \node[kstate,fill=lightgray!50] (P4) at (  6,0) {$\relint{F}$};
    \node[kstate,fill=lightgray!50] (P5) at (  5,0) {$\relint{E}$};

    \node[kstate,fill=red!50] (E0) at (-1,1) {$\relint{AB}$};
    \node[kstate,fill=red!50] (E1) at ( 2,1) {$\relint{BD}$};
    \node[kstate,fill=red!50] (E2) at ( 1,1) {$\relint{BC}$};
    \node[kstate,fill=red!50  ] (E3) at ( 0,1) {$\relint{AC}$};
    \node[kstate,fill=red!50  ] (E4) at ( 3,1) {$\relint{CD}$};
	\node[kstate,fill=lightgray!50] (E5) at ( 6,1) {$\relint{DF}$};
	\node[kstate,fill=lightgray!50] (E6) at ( 5,1) {$\relint{DE}$};
	\node[kstate,fill=lightgray!50] (E7) at ( 4,1) {$\relint{CE}$};
	\node[kstate,fill=lightgray!50] (E8) at ( 7,1) {$\relint{EF}$};

    \node[kstate,fill=red!50] (T0) at ( 2,2) {$\relint{BCD}$};
    \node[kstate,fill=red!50] (T1) at ( 0,2) {$\relint{ABC}$};
    \node[kstate,fill=lightgray!50] (T2) at ( 6,2) {$\relint{DEF}$};
    \node[kstate,fill=green!50] (T3) at ( 4,2) {$\relint{CDE}$};

    \draw (P0) to (E0);
    \draw (P0) to (E1);
    \draw (P0) to (E2);

    \draw (P1) to (E0);
    \draw (P1) to (E3);

    \draw (P2) to (E1);
    \draw (P2) to (E4);
    \draw (P2) to (E5);
    \draw (P2) to (E6);

    \draw (P3) to (E2);
    \draw (P3) to (E3);
    \draw (P3) to (E4);
    \draw (P3) to (E7);

    \draw (P4) to (E5);
    \draw (P4) to (E8);

    \draw (P5) to (E6);
    \draw (P5) to (E7);
    \draw (P5) to (E8);
    \draw (E0) to (T1);
    \draw (E2) to (T0);
    \draw (T1) to (E2);
    
    \draw (E1) to (T0);   
    \draw (E3) to (T1);    
    \draw (E4) to (T0);
    \draw (E4) to (T3);
    \draw (E5) to (T2);
    \draw (E6) to (T2);
	\draw (E6) to (T3);
	\draw (E7) to (T3);
	\draw (E8) to (T2);

\end{tikzpicture}
}
}
\caption{A polyhedral model $\calP$
  (\ref{subfig:PolyhedronNoPathCompressed}) with its cells
  (\ref{subfig:PolyhedronNoPathCellsCompressed}) and the Hasse diagram
  of the related cell poset
  (\ref{subfig:PolyhedronNoPathPosetCompressed}).} 
\label{fig:PolyhedronNoPathCompressed}
\end{figure}

A \emph{polyhedral model} is a pair $\calP = (|K|,\peval_{\calP})$
where $\peval_{\calP}: \ap \to \pws{|K|}$ maps every proposition letter
$p \in \ap$ to the set of points of $|K|$ satisfying~$p$. It is
required that, for all $p \in \ap$, $\peval_{\calP}(p)$ is always a
union of cells in~$\relint{K}$.  Similarly, a poset model
$\calF = (W,\preccurlyeq,\peval_{\calF})$ is a poset $(W,\preccurlyeq)$
equipped with a valuation function
$\peval_{\calF} : \ap \to \pws{W}\!$. Given a polyhedral model
$\calP = (|K|,\peval_{\calP})$, we say that
$\calF = (\relint{K},\preccurlyeq,\peval_{\calF})$ is the \emph{cell
  poset model} of~$\calP$ if and only if $(\relint{K},\preccurlyeq)$
is the cell poset of~$K$ and, for all $\relint{\sigma}\in \relint{K}$,
we have: $\relint{\sigma} \in \peval_{\calF}(p)$ if and only if
$\relint{\sigma} \subseteq \peval_{\calP}(p)$.
  For all $x \in |K|$, we let $\map(x)$ denote the unique
  cell~$\relint{\sigma}$ such that $x \in \relint{\sigma}$. Note that
  $\map(x)$ is well defined, since $\relint{K}$ is a partition of $|K|$ and that
  $\map: |K| \to \relint{K}$ is a continuous function~\cite[Corollary
  3.4]{BMMP2018}. With slight overloading, we let
  $\map(\calP)$~denote the cell poset model of~$\calP$. Finally, note
  that poset models are a subclass of Kripke models. In the following,
  when we say that $\calF$ is a cell poset model, we mean that there
  exists a polyhedral model $\calP$ such that $\calF = \map(\calP)$.
Fig.~\ref{fig:PolyhedronNoPathCompressed} shows a polyhedral
model. There are three proposition letters, $\mathbf{red}$,
$\mathbf{green}$, and $\mathbf{gray}$, shown by different colours
(\ref{subfig:PolyhedronNoPathCompressed}). The model is ``unpacked''
into its cells in
Fig.~\ref{subfig:PolyhedronNoPathCellsCompressed}. The latter are
collected in the cell poset model, whose Hasse diagram is shown in
Fig.~\ref{subfig:PolyhedronNoPathPosetCompressed}.
\subsubsection{Paths.}
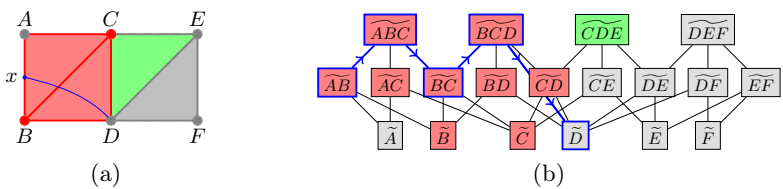
\begin{figure}[t]
\begin{center}
\subfloat[]{\label{subfig:PolyhedronWithPath}
\resizebox{1.2in}{!}{
\begin{tikzpicture}[scale=1.3,label distance=-2pt]
	    \tikzstyle{point}=[circle,draw=black,fill=white,inner sep=0pt,minimum width=4pt,minimum height=4pt]
	    \node (p0)[point,draw=red,label={270:$B$}] at (0,0) {};
	    	\filldraw [red] (p0) circle (1.25pt);
%	    \node (p1)[point,draw=red,label={ 90:$A$}] at (0,1) {};
	    \node (p1)[point,draw=gray,label={ 90:$A$}] at (0,1) {};
%	    	\filldraw [red] (p1) circle (1.25pt);
	    	\filldraw [gray] (p1) circle (1.25pt);
	    \node (p2)[point,draw=gray,label={270:$D$}] at (1,0) {};
	    \node (p3)[point,draw=red,label={ 90:$C$}] at (1,1) {};
	    	\filldraw [red] (p3) circle (1.25pt);
	    \node (p4)[point,draw=gray,label={270:$F$}] at (2,0) {};
	    \node (p5)[point,draw=gray,label={ 90:$E$}] at (2,1) {};

	    \draw [red   ,thick](p0) -- (p1);
	    \draw [red   ,thick](p0) -- (p2);
	    \draw [red   ,thick](p0) -- (p3);
	    \draw [red   ,thick](p1) -- (p3);
	    \draw [red   ,thick](p2) -- (p3);	    
    \draw [dashed      ](p2) -- (p4);
    \draw [dashed      ](p2) -- (p5);
    \draw [dashed      ](p3) -- (p5);
    \draw [dashed      ](p4) -- (p5);
    \draw [gray,thick](p2) -- (p4);
    \draw [gray,thick](p4) -- (p5);
    \draw [gray,thick](p2) -- (p5);
    \draw [gray,thick](p3) -- (p5);
	        
	    \begin{scope}[on background layer]
	    \fill [fill=red!50  ](p0.center) -- (p1.center) -- (p3.center);
	    \fill [fill=red!50  ](p0.center) -- (p3.center) -- (p2.center);
	    \fill [fill=green!50](p2.center) -- (p3.center) -- (p5.center);
            \fill [fill=gray!50](p2.center) -- (p4.center) -- (p5.center);    	    
            \end{scope}

	    \node at (-.15,.5) {$x$};
	    \fill[blue] (0,.5) circle (.7pt);
	    \draw [blue] plot [smooth,tension=1] coordinates { (0,.5) (.6,.3) (1,0)};
	    \fill[blue] (1,0) circle (.7pt);
    \filldraw [gray] (p2) circle (1.25pt);
    \filldraw [gray] (p4) circle (1.25pt);
    \filldraw [gray] (p5) circle (1.25pt); 
	\end{tikzpicture}
}
	}\quad\quad\quad
\subfloat[]{\label{subfig:PosetWithPath}
\resizebox{2.5in}{!}{
\begin{tikzpicture}[scale=0.8, every node/.style={transform shape}]
    \tikzstyle{kstate}=[rectangle,draw=black,fill=white]
    \tikzset{->-/.style={decoration={
		markings,
		mark=at position #1 with {\arrow{>}}},postaction={decorate}}}
    
    \node[kstate,fill=red!50  ] (P0) at (  1,0) {$\relint{B}$};
    \node[kstate,fill=lightgray!50  ] (P1) at (  0,0) {$\relint{A}$};
    \node[kstate,fill=lightgray!50,draw=blue,thick] (P2) at (3.5,0) {$\relint{D}$};
    \node[kstate,fill=red!50  ] (P3) at (2.5,0) {$\relint{C}$};
    \node[kstate,fill=lightgray!50] (P4) at (  6,0) {$\relint{F}$};
    \node[kstate,fill=lightgray!50] (P5) at (  5,0) {$\relint{E}$};

    \node[kstate,fill=red!50  ,draw=blue,thick] (E0) at (-1,1) {$\relint{AB}$};
    \node[kstate,fill=red!50  ] (E1) at ( 2,1) {$\relint{BD}$};
    \node[kstate,fill=red!50  ,draw=blue,thick] (E2) at ( 1,1) {$\relint{BC}$};
    \node[kstate,fill=red!50  ] (E3) at ( 0,1) {$\relint{AC}$};
    \node[kstate,fill=red!50  ] (E4) at ( 3,1) {$\relint{CD}$};
	\node[kstate,fill=lightgray!50] (E5) at ( 6,1) {$\relint{DF}$};
	\node[kstate,fill=lightgray!50] (E6) at ( 5,1) {$\relint{DE}$};
	\node[kstate,fill=lightgray!50] (E7) at ( 4,1) {$\relint{CE}$};
	\node[kstate,fill=lightgray!50] (E8) at ( 7,1) {$\relint{EF}$};

    \node[kstate,fill=red!50  ,draw=blue,thick] (T0) at ( 2,2) {$\relint{BCD}$};
    \node[kstate,fill=red!50  ,draw=blue,thick] (T1) at ( 0,2) {$\relint{ABC}$};
    \node[kstate,fill=lightgray!50] (T2) at ( 6,2) {$\relint{DEF}$};
    \node[kstate,fill=green!50] (T3) at ( 4,2) {$\relint{CDE}$};

    \draw (P0) to (E0);
    \draw (P0) to (E1);
    \draw (P0) to (E2);

    \draw (P1) to (E0);
    \draw (P1) to (E3);

    \draw (P2) to (E1);
    \draw (P2) to (E4);
    \draw (P2) to (E5);
    \draw (P2) to (E6);

    \draw (P3) to (E2);
    \draw (P3) to (E3);
    \draw (P3) to (E4);
    \draw (P3) to (E7);

    \draw (P4) to (E5);
    \draw (P4) to (E8);

    \draw (P5) to (E6);
    \draw (P5) to (E7);
    \draw (P5) to (E8);

    \draw[blue,thick,->-=.5] (E0) to (T1);

    \draw (E1) to (T0);
    
    \draw[blue,thick,->-=.5] (E2) to (T0);
    
    \draw[blue,thick,->-=.5] (T1) to (E2);

    \draw (E3) to (T1);
    
    \draw (E4) to (T0);
    \draw (E4) to (T3);

	\draw (E5) to (T2);

	\draw (E6) to (T2);
	\draw (E6) to (T3);

	\draw (E7) to (T3);

	\draw (E8) to (T2);

	\begin{scope}[on background layer]
		\draw[blue,thick,->-=.2] (T0) to (P2);
		\draw[blue,thick,->-=.8] (T0) to (P2);
	\end{scope}
\end{tikzpicture}
}
}
\end{center}
\caption{(\ref{subfig:PolyhedronWithPath}) A topological path from a
  point $x$ to vertex $D$ in the polyhedral model $\calP$ of
  Figure~\ref{subfig:PolyhedronNoPathCompressed}. (\ref{subfig:PosetWithPath})
  The corresponding \plm-path (in blue) in the Hasse diagram of the
  cell poset model $\map(\calP)$.}
\label{fig:poset}
\end{figure}

In a topological space $(X,\tau)$, a \emph{topological path} from
$x\in X$ is a total, continuous function $\pi : [0,1] \to X$ such that
$\pi(0)=x$.  We call $\pi(0)$ and~$\pi(1)$ the \emph{starting point}
and \emph{ending point} of~$\pi$, respectively, while $\pi(r)$~is an
\emph{intermediate point} of~$\pi$, for all $r \in
(0,1)$. Fig.~\ref{subfig:PolyhedronWithPath} shows a path from a
point~$x$ in the open segment~$\relint{AB}$ in the polyhedral model of
Fig~\ref{subfig:PolyhedronNoPathCompressed}.
Topological paths relevant for our work 
are represented in cell posets by so-called \plm-paths, a subclass of
undirected paths~\cite{Be+22}. For technical reasons\footnote{We are
  interested in model-checking structures resulting from the
  minimisation, via bisimilarity, of cell poset models, and such
  structures are often just (reflexive) Kripke models rather than
  poset models.}
in this paper we extend the definition given in~\cite{Be+22} to
general Kripke frames. These are sets $W$ of elements, usually called nodes or worlds,
together with a preordering relation $R$, called its accessibility relation.
Given a Kripke frame $(W,R)$, an \emph{undirected path} of length
$\ell \in \nats$ from~$w$ is a total function $\pi : [0;\ell] \to W$
such that $\pi(0) = x$ and, for all $i \in [0;\ell)$,
$\dircnv{R}(\pi(i),\pi(i+1))$. The \emph{starting point} and
\emph{ending point} are $\pi(0)$ and~$\pi(\ell)$, respectively, while
$\pi(i)$ is an intermediate point, for all $i \in (0;\ell)$. For an
undirected path~$\pi$ of length~$\ell$ we often use the sequence
notation $(w_i)_{i=0}^{\ell}$ where $w_i = \pi(i)$ for $i \in [0;\ell]$.  
An undirected path $\pi : [0;\ell] \to W$ is a \emph{\plm-path} if and
only if $\ell\geq 2$, $R(\pi(0),\pi(1))$ and
$\cnv{R}(\pi(\ell-1),\pi(\ell))$.
The \plm-path
$(\relint{AB},\relint{ABC},\relint{BC},\relint{BCD},\relint{D})$,
drawn in blue in Fig.~\ref{subfig:PosetWithPath}, faithfully
represents the path from~$x$ shown in
Fig.~\ref{subfig:PolyhedronWithPath}. 

\subsubsection{Spatial logic.} In~\cite{Be+22}, the spatial logic \slcsG, a version of \slcs{} for polyhedral models,
has been presented that consists of predicate letters, negation,
conjunction, and the single modal operator~$\gamma$, expressing
conditional reachability. 

For $p \in \ap$ the syntax of the logic is the following:
%
%\begin{equation}\label{def:syntax-minimal}
$$
\form  ::=   p  \msep  \lneg \, \form  \msep  \form \, \land \, \form  \msep \gamma(\form,\form)% \msep \cnvnear \form
$$
%\end{equation}

The satisfaction relation for
$\gamma(\form_1, \form_2)$, for a polyhedral model
$\calP = (|K|,\peval_{\calP})$ and
$x\in |K|$, as defined in~\cite{Be+22}, is recalled below:\\[0.5em]
$
\begin{array}{l c l}
\calP, x \models \gamma(\form_1,\form_2) & \Leftrightarrow &
\mbox{a } \mbox{topological path } \pi: [0,1] \to |K| \mbox{ exists such that } \pi(0)=x,\\&&
\calP, \pi(1) \models \form_2, \mbox{and }
\calP, \pi(r) \models \form_1 \mbox{ for all } r\in \!\!(0,\!1).
\end{array}
$\\[0.5em]
We also recall the interpretation of \slcsG{} on poset models. The
satisfaction relation for $\gamma(\form_1, \form_2)$, for a poset model
$\calF=(W,\preccurlyeq,\peval_{\calF})$ and
$w\in W$, is as follows:\\[0.5em]
$
\begin{array}{l c l}
\calF, w \models \gamma(\form_1,\form_2) & \Leftrightarrow &
\mbox{a } \mbox{\plm-path } \pi: [0;\ell] \to W \mbox{ exists such that } \pi(0)=w,\\&&
\calF, \pi(\ell) \models \form_2, \mbox{and }
\calF, \pi(i) \models \form_1 \mbox{ for all } i\in \!\!(0;\!\ell).
\end{array}
$\\[0.5em]
The near operator $\dirnear$, corresponding to the classical closure operator, can be defined as a derived operator: $\dirnear \form = \gamma(\form,\ltrue)$. In the context of posets $w \models \dirnear \form$ means that there exists a \plm-path starting from $w$ leading in one step going up to a cell that satisfies $\form$.

\subsubsection{Bisimulation based model minimisation.} In~\cite{Be+22} it has also been shown that, for all $x \in |K|$ and
\slcsG{} formulas~$\form$, we have: $\calP,x \models \form$ if and
only if $\map(\calP),\map(x) \models \form$.  In addition,
\emph{simplicial bisimilarity}, a novel notion of bisimilarity for
polyhedral models, has been defined that uses a subclass of
topological paths and it has been shown to enjoy the classical
Hennessy-Milner property: two points $x_1,x_2 \in |K|$ are simplicial
bisimilar, written $x_1 \sibis^{\calP} x_2$, if and only if they
satisfy the same \slcsG{} formulas, i.e. they are equivalent with
respect to the logic \slcsG{}, written $x_1 \slcsGeq^{\calP} x_2$.

The result has been extended to \emph{\plm-bisimilarity} on finite
poset models, a notion of bisimilarity based on \plm-paths:
$w_1,w_2 \in W$ are \plm-bisimilar, written $x_1 \plmbis^{\calF} x_2$,
if and only if they satisfy the same \slcsG{} formulas, i.e.
$x_1 \slcsGeq^{\calF} x_2$ (see~\cite{Ci+23c} for details).  In
summary, we have:
\[
x_1 \sibis^{\calP} x_2 \mbox{ iff }
x_1 \slcsGeq^{\calP} x_2 \mbox{ iff }
\map(x_1) \slcsGeq^{\map(\calP)} \map(x_2) \mbox{ iff }
\map(x_1)\plmbis^{\map(\calP)} \map(x_2).
\]

In~\cite{Be+24} we showed a similar result for a weaker logic, namely \slcsE{,} where $\gamma(\form_1,\form_2)$ is replaced by $\eta(\form_1,\form_2)$, which also requires the satisfaction of $\form_1$. In other words, requiring $\eta(\form_1,\form_2)$ is the same as requiring  $\form_1 \land \gamma(\form_1,\form_2)$. Weaker notions of bisimilarity, $\wsibis$ and $\wplmbis$, have been introduced and the following has been proven:
\[
x_1 \wsibis^{\calP} x_2 \mbox{ iff }
x_1 \slcsEeq^{\calP} x_2 \mbox{ iff }
\map(x_1) \slcsEeq^{\map(\calP)} \map(x_2) \mbox{ iff }
\map(x_1)\wplmbis^{\map(\calP)} \map(x_2).
\]

In~\cite{Be+24b} we presented an effective toolchain for model minimisation modulo this weaker bisimilarity, and thus also modulo \slcsE{.}
Let us provide a small example to illustrate the spatial bisimulation. With reference to
Figure~\ref{subfig:PolyhedronNoPathCompressed}, we have that no red
point, call it~$y$, in the open segment~$CD$ is simplicial bisimilar
to the red point~$C$. In fact, although both $y$ and~$C$ satisfy
$\gamma(\mathbf{green}, \ltrue)$, we have that $C$~satisfies also
$\gamma(\mathbf{gray}, \ltrue)$, which is not the case for~$y$.
Similarly, with reference to
Figure~\ref{subfig:PolyhedronNoPathPosetCompressed}, cell~$\relint{C}$
satisfies $\gamma(\mathbf{gray}, \ltrue)$, which is not satisfied
by~$\relint{CD}$.

\subsubsection{Extension of \slcsG{.}} In this work we will add a further operator, converse near ($\cnvnear$), that expresses the converse of $\dirnear$. %The latter corresponds to classical topological closure. 
As we have seen, $\dirnear$ can be defined as a derived operator in terms of $\gamma$, but this is not the case for $\cnvnear$.\footnote{An investigation on a `converse' of the $\gamma$ modality,  say $\stackrel{\leftarrow}{\gamma}$, from which $\cnvnear$ could be derived, is beyond the scope of the present paper.} 
The satisfaction relation for
$\cnvnear \form$, for a polyhedral model
$\calP = (|K|,\peval_{\calP})$ and
$x\in |K|$, is defined as:\\[0.5em]
$
\begin{array}{l c l}
\calP, x \models \cnvnear \form & \Leftrightarrow &
\mbox{ exists } x' \in \closure_T(\map(x)) \mbox{ s.t. } x' \models \form
\end{array}
$\\[0.5em]
The satisfaction relation for  $\cnvnear$, for a poset model
$\calF=(W,\preccurlyeq,\peval_{\calF})$ and
$w\in W$, is defined as:\\[0.5em]
$
\begin{array}{l c l}
\calF, w \models \cnvnear \form & \Leftrightarrow &
\mbox{ exists } w' \mbox{ s.t. } w' \preccurlyeq w \mbox{ and } w' \models \form
\end{array}
$\\[0.5em]
We leave the treatment of spatial bisimulation and minimisation for the extended logic to future work.

\subsubsection{Polyhedral model checking and visualisation.} As we have seen above, \slcsG\ properties of polyhedra can be transformed into equivalent properties on cell poset models. This result is exploited in the polyhedra model checker \polylogica\ and in the tool  \polyvisualizer\ to visualise the model checking results. \polylogica\ is a global model checker that checks a given formula for all cells of the model at once. The algorithm takes as input a Kripke poset model $\map(\calP)$ of polyhedron $\calP$ and an \slcsG\ formula $\phi$. The output is the satisfaction set $\mbox{\sf Sat}(\phi) = \ZET{\relint{\sigma} \in \relint{K}}{\map(\calP),\relint{\sigma} \models \phi}$ of nodes in the poset model $\map(\calP)$ that correspond to the set of cells of $\calP$ that satisfy formula $\phi$. The satisfaction set {\sf Sat} is defined recursively on the structure of \slcsG\ formula~\cite{Be+22}. 

The actual input language of \polylogica{,} called \imgql\ (image query language), is an extended set of operations that has \slcsG\ as its core language. The \polylogica\ model checker\footnote{\polylogica\ 0.4 and \polyvisualizer\ are available in the branch {\sf polyhedra} of the main \voxlogica\ repository, see {\sf https://github.com/vincenzoml/VoxLogicA.}} is written in {\sf FSharp}. The model checking results of \polylogica\ are generated as a {\sf json} file. This file contains the name of the property as defined in the \imgql\ specification, followed by a list of {\sf true} and {\sf false} following the order of the definition of the cells in the poset model. Cells for which the corresponding position in the {\sf json} list is {\sf true} satisfy the property, cells corresponding to {\sf false} do not satisfy it. The result file is used by the \polyvisualizer\ tool to present the model checking results for the polyhedron in a visual way. Cells that satisfy the property are highlighted in their original colour, those that do not satisfy the property are shown in a semi-transparent way. We will see examples of this use in the next section where model checking results are shown as screenshots of the 3D images produced by the \polyvisualizer\ tool (see for example Fig.~\ref{fig:3Dmaze}b showing all cells satisfying the formula $B \lor W$, where $B$ stands for the atomic proposition {\sf black} and $W$ for {\sf white}). Details on the polyhedra model checking algorithm and the visualiser can be found in~\cite{Be+22}.  

\section{Polyhedral Model Checking Examples}\label{sec:examples}

We illustrate the potential of polyhedral model checking for three case studies, each illustrating different uses. The first is a 3D maze model. The second is a model of a tree-shaped coral. The third is a model of a simple aircraft. The original models of the latter two were provided in the Wavefront .obj format. The latter format is widely used in computer graphics. We have developed dedicated python scripts to convert Wavefront .obj images into an input format suitable for polyhedra model checking with \polylogica\ and for viewing with the \polyvisualizer{.} An example of such a script can be found in the Appendix.

\subsection{Analysing Reachability in a 3D Maze Model}
\begin{figure}[h!]
\centering
	\subfloat[]{\label{subfig:full}
		\includegraphics[valign=c,height=6em]{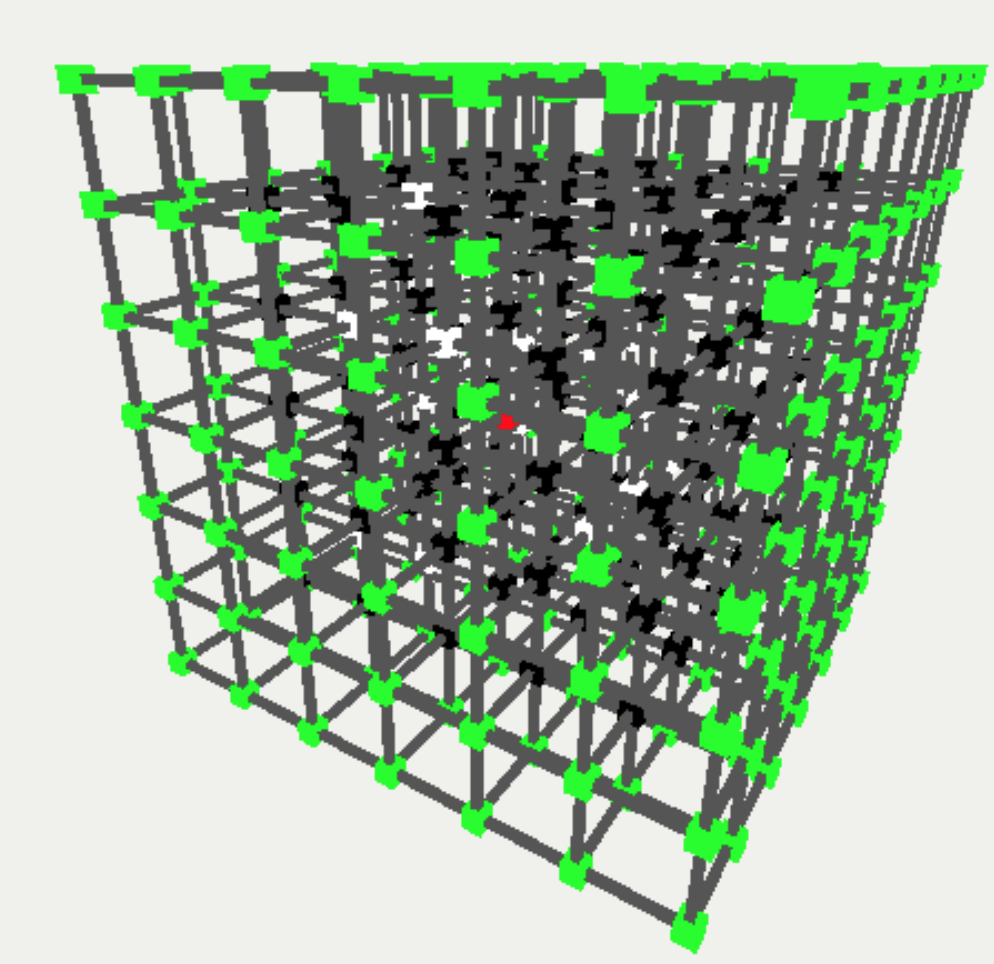}
	}
	\subfloat[]{\label{subfig:black_or_white_rooms}
		\includegraphics[valign=c,height=6em]{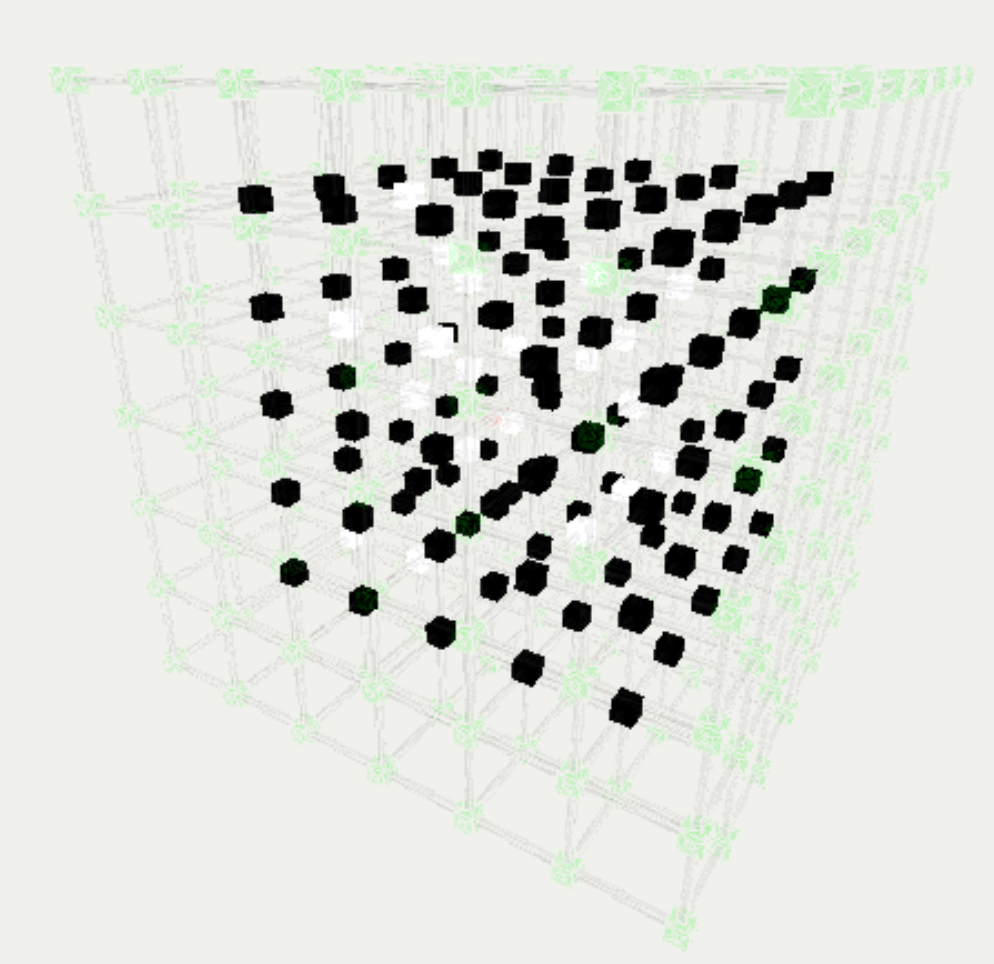}
	}
	\subfloat[]{\label{subfig:red_rooms}
		\includegraphics[valign=c,height=6em]{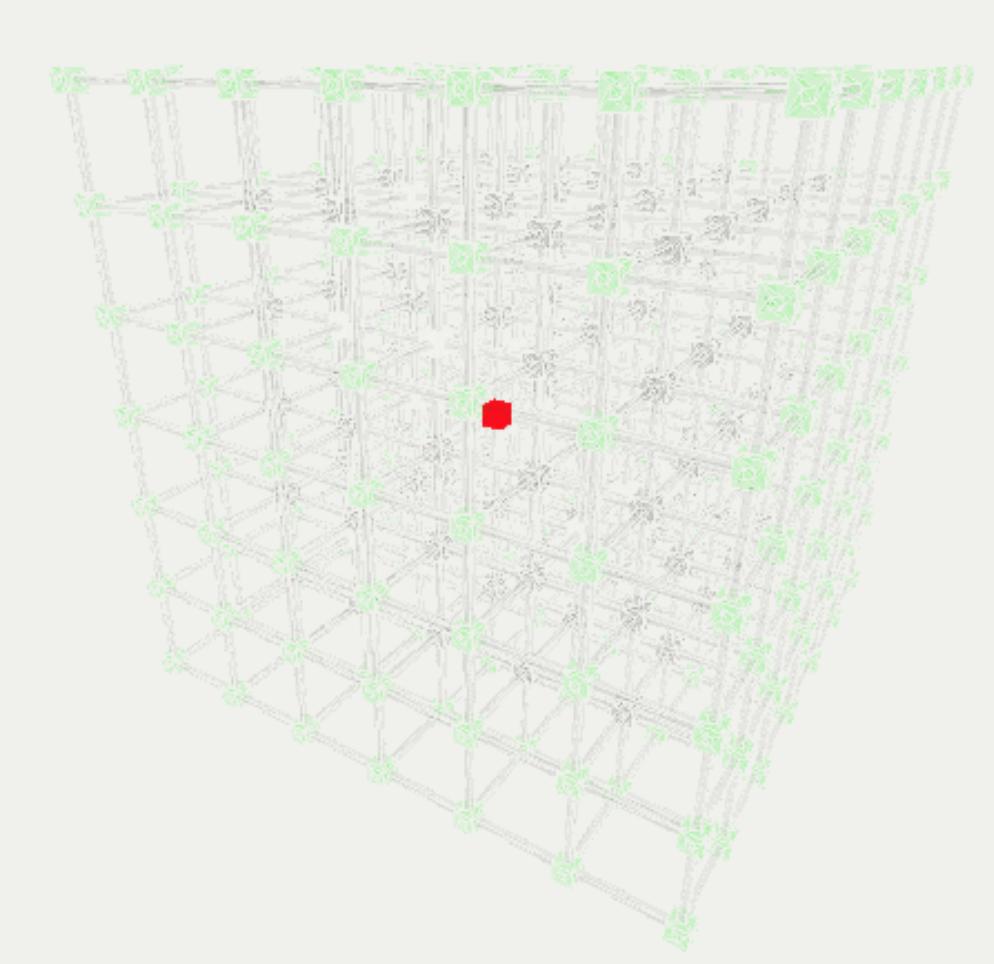}
	}
\caption{ 3D maze~(\ref{subfig:full}), black and white~(\ref{subfig:black_or_white_rooms}) and red rooms~(\ref{subfig:red_rooms}) in the 3D maze.}\label{fig:3Dmaze}
\end{figure}

As a first example we consider a volumetric mesh of a 3D maze which we recall briefly~\cite{Be+22}. Its purpose is to illustrate how \polylogica\ can be used to reason about cell colours of an object, where the colours are treated as atomic propositions. The maze, shown in Figure~\ref{subfig:full}, is a square structure that consists of 7 by 7 by 7 `rooms' that are connected by `corridors'. Each room and corridor is a polyhedron composed, in turn, by a number of cells such as vertices, segments, triangles and tetrahedra.

The rooms at the outmost sides of the cube are green. Inside the cube we find black and white rooms (see Fig.~\ref{subfig:black_or_white_rooms}) and one red room (see Fig.~\ref{subfig:red_rooms}). The corridors are all dark grey. The images in Fig.~\ref{fig:3Dmaze} have all been obtained by polyhedral model checking with \polylogica{.} In particular, Fig.~\ref{subfig:black_or_white_rooms} shows all cells that satisfy the formula $B \lor W$, where $B$ is the atomic proposition for black cells and $W$ the atomic proposition for white cells. The cells that satisfy the formula are shown highlighted in their original colour. The cells that do not satisfy the formula are shown in a pale version of their original colour. The cells satisfying the atomic proposition $R$ (red cells) are shown in a similar way in Fig.~\ref{subfig:red_rooms}.

Fig.~\ref{fig:SLCSQ1Q2Q3} shows some \slcsG\ formulas that characterise various kinds of corridors, for example, corridors connecting white rooms on both sides is defined using the reachability operator $\gamma$, where $\gamma(C,W)$ requires the corridor to connect to a white room and $\neg\gamma(C, G \lor B \lor R)$ makes sure it does not connect to a room of any other colour.

Formulas $Q1$, $Q2$ and $Q3$ express a few less basic \slcsG\ properties. $Q1$: White rooms and their connecting corridors from which a green room can be reached not passing by black rooms, including the green room that is reached (\texttt{connWG}); $Q2$: White rooms and their connecting corridors from which both a red and a green room can be reached not passing by black rooms (\texttt{connRWG}); $Q3$: White rooms and their connecting corridors with no path to green rooms  (\texttt{whiteNoGreen} or equivalently \texttt{whiteSblack}). The model checking results of $Q1$, $Q2$ and $Q3$ are shown in Fig.~\ref{fig:3DmazeMC}, again by highlighting the cells that satisfy the specific formula in their original colour.

\begin{figure}[t!]
\centering
\fbox{
$
\begin{array}{l c l}
\mbox{\sf corridorWW} &\equiv & \gamma(C,W) \land \neg\gamma(C, G \lor B \lor R)\\
\mbox{\sf corridorWG} &\equiv & \gamma(C,W) \land \gamma(C, G)\\
\mbox{\sf corridorWR} &\equiv & \gamma(C,W) \land \gamma(C, R)\\
\mbox{\sf corridorWB} &\equiv & \gamma(C,W) \land \gamma(C, B)\\
\mbox{\sf whiteToGreen} &\equiv & \gamma((W \lor \mbox{\sf corridorWW} \lor \mbox{\sf corridorWG}),G)\\
\mbox{\sf Q1} &\equiv &  \mbox{\sf whiteToGreen} \lor \gamma(G,\mbox{\sf whiteToGreen})\\
\mbox{\sf Q2} &\equiv & \gamma((\mbox{\sf Q1} \lor \mbox{\sf corridorWR}), R) \lor \gamma((R \lor \mbox{\sf corridorWR}),\mbox{\sf Q1})\\
\mbox{\sf Q3} &\equiv & (W \lor \mbox{\sf corridorWW}) \land \neg \mbox{\sf whiteToGreen}.
\end{array}
$
}
\caption{\slcsG\ formulas expressing properties {\sf Q1}, {\sf Q2} and {\sf Q3}; atomic proposition letters $G,W,B,R,C$ are assumed given and their meaning is the obvious one ($C$ for ``corridor'', $G$ for green and similarly for the other colours).\label{fig:SLCSQ1Q2Q3}}
\end{figure}

\begin{figure}[h!]
\centering
	\subfloat[Q1]{\label{subfig:connectionWhiteGreen}
		\includegraphics[valign=c,height=6em]{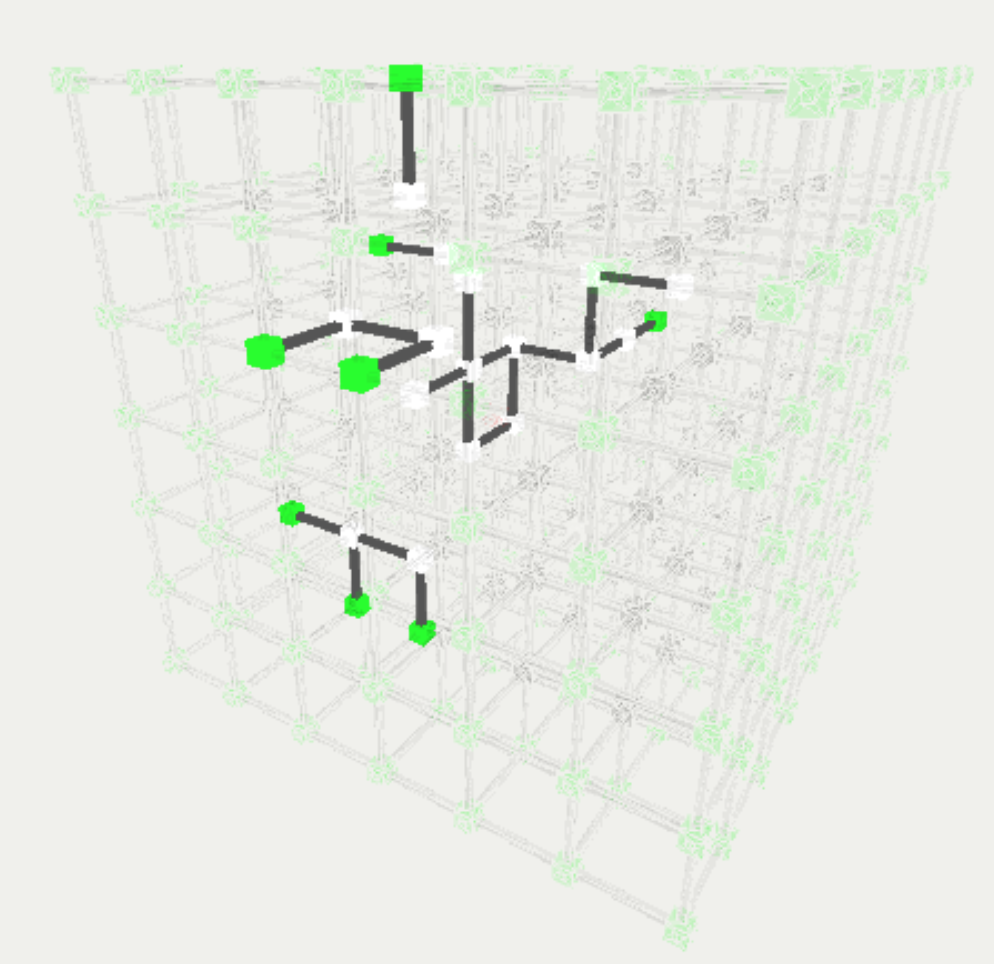}
	}
	\subfloat[Q2]{\label{subfig:whiteConnectsRedGreen}
		\includegraphics[valign=c,height=6em]{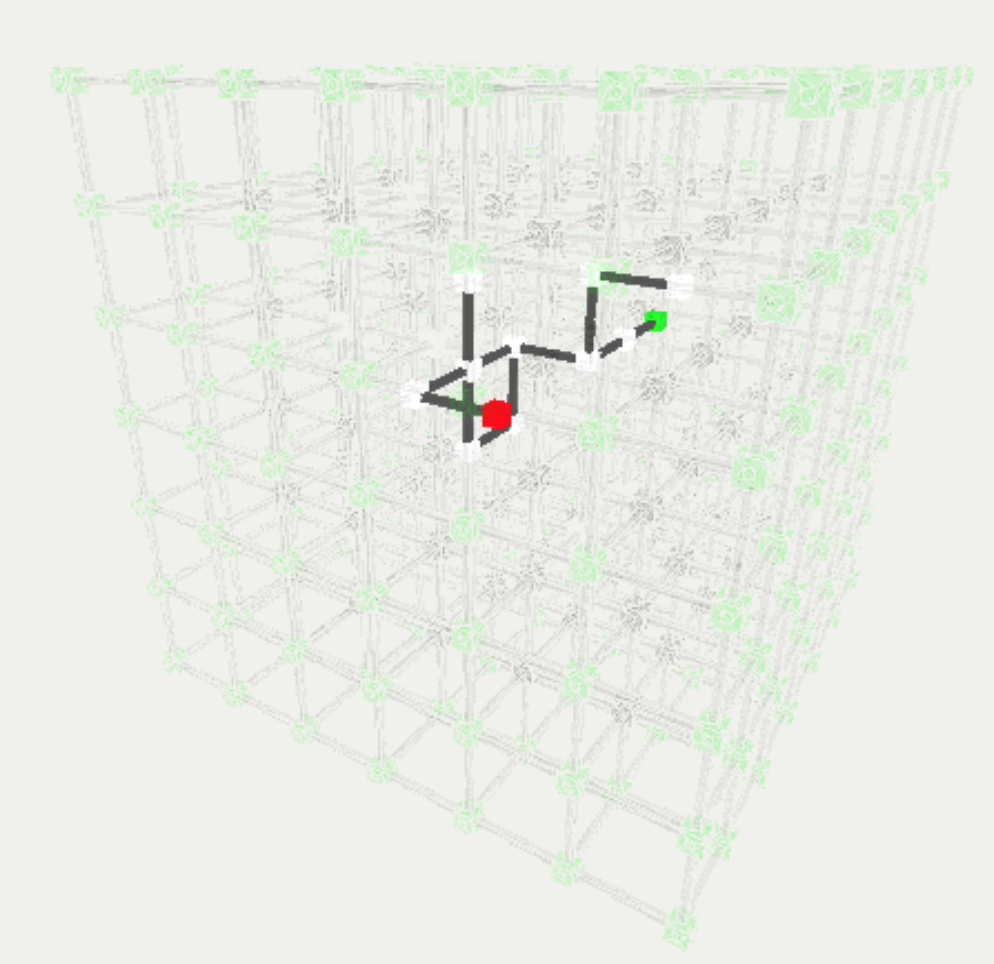}
	}
	\subfloat[Q3]{\label{subfig:no_exit_rooms}
		\includegraphics[valign=c,height=6em]{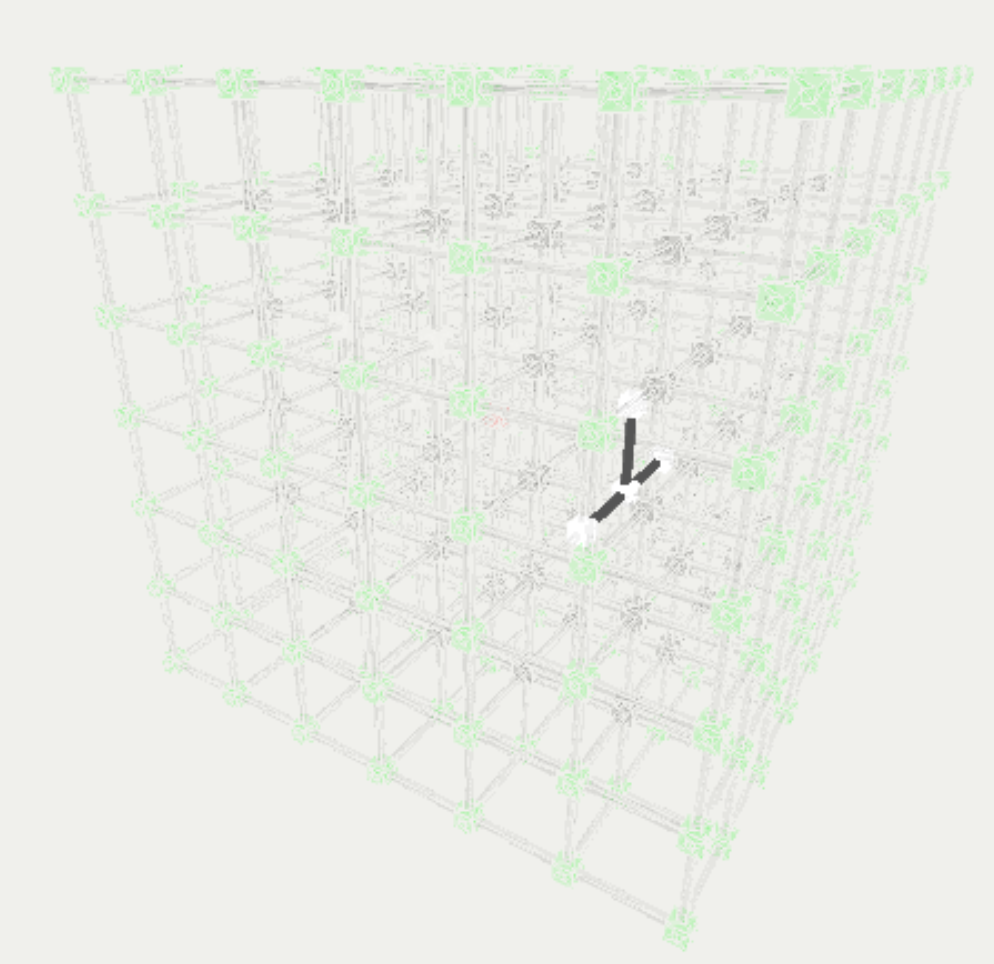} 
	}
\caption{ Spatial model checking results of the properties in Figure~\ref{fig:SLCSQ1Q2Q3} for the 3D maze of Figure~\ref{fig:3Dmaze}. }\label{fig:3DmazeMC}
\end{figure}

The valid paths through the maze should only pass by white rooms (and related corridors) to reach a green room without passing by black rooms or corridors that connect to black rooms.
The 3D maze model consists of 147,245 cells and the approximate time to analyse all properties in Fig.~\ref{fig:3DmazeMC} together amounts to about 6 seconds\footnote{All experiments in this paper were performed with \polylogica\ 0.4 on a Mac M2 pro with 12 cores and 32GB of memory.} (of which 5.5 seconds for loading and parsing the model file). The full \imgql\ specification can be found in Appendix~\ref{app:maze}.

\subsection{Analysing the Branching Hierarchy of a Coral Model}

The second example is used to illustrate a semi-automatic approach to find the various branches in a tree-shape coral (see~\cite{Andriaccio2022,Andriaccio+2024}). In this example, \polylogica\ is used in combination with the tool MeshLab. The latter is used to perform some initial manual annotations of the model and then saved in the Wavefront .obj format. It is then converted, with the help of a dedicated python script (see Annex~\ref{sec:convert}), into an input format that is suitable for further analysis using \polylogica{.} 

In the following, we'll show that the results of \polylogica\ model checking can be used to enrich the model itself with additional atomic propositions and then used again for further model checking or for model minimisation. This novel way of enriching a model with previously obtained spatial model checking results is very interesting and has many advantages. First of all, it can be used to keep the formulas of interest much shorter, resulting in reduced model checking times. This is important as models tend to be very large. Second, enriching/changing models this way makes it possible to study their core structure through minimisation of the model. Third, such minimised models may, in their turn, be used to speed up further model checking.

 \begin{figure}[t!]
\centering
\subfloat[\label{fig:corala}]{\includegraphics[width=0.45\textwidth]{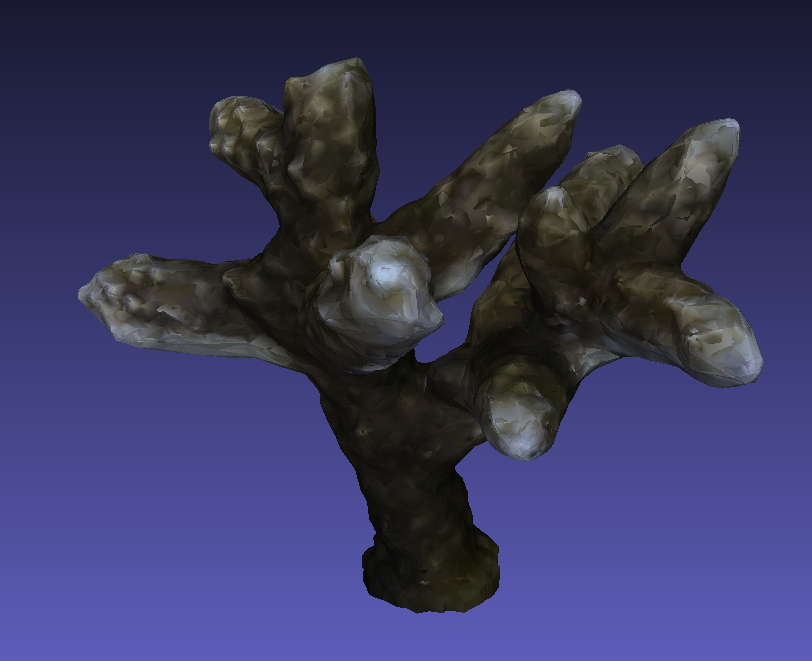}}\quad\quad\quad
\subfloat[\label{fig:coralb}]{\includegraphics[width=0.45\textwidth]{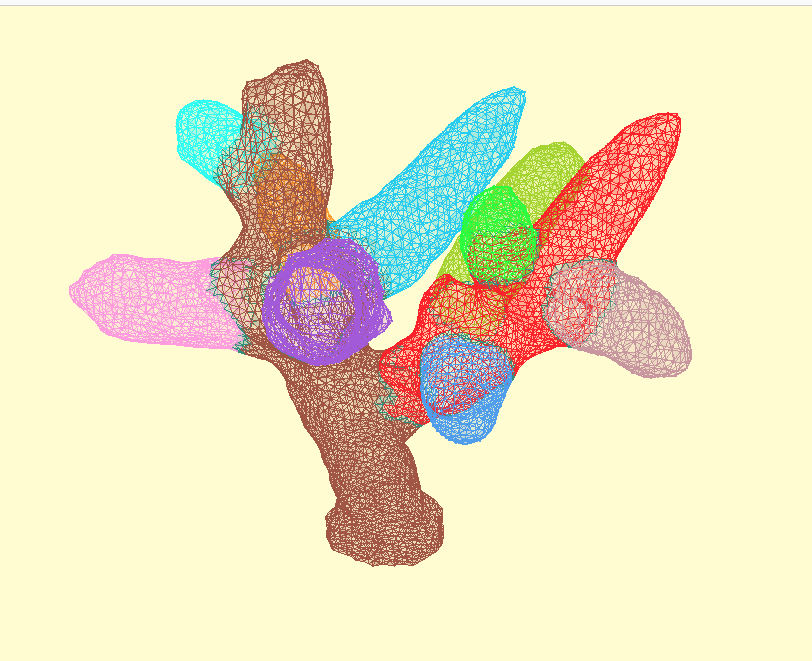}}
\caption{(a) Tree-shape coral in MeshLab; (b) Coral with model checking results for finding branches.}\label{fig:coral}
\end{figure}

\begin{figure}[t!]
\centering
\subfloat[\label{fig:coralBa}]{\includegraphics[width=0.45\textwidth]{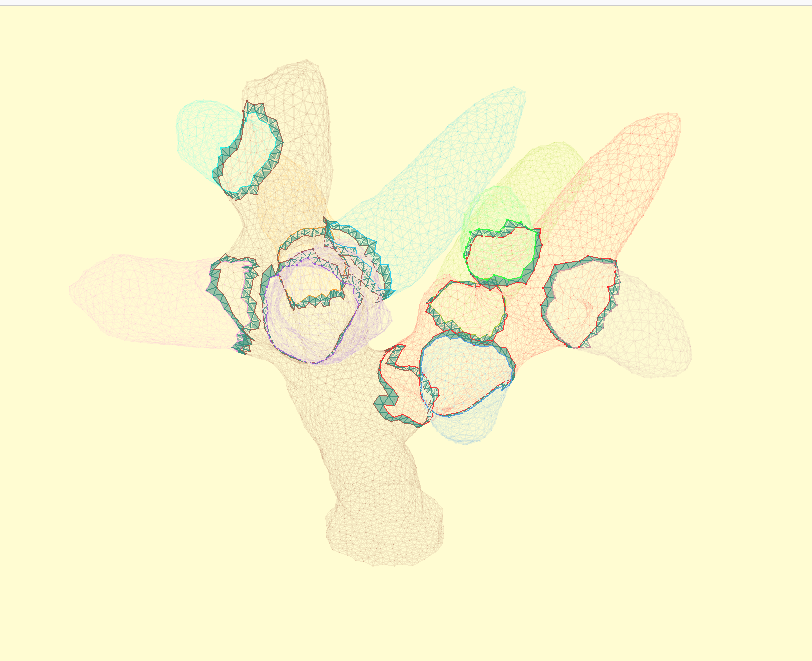}}\quad\quad\quad
\subfloat[\label{fig:coralBb}]{\includegraphics[width=0.45\textwidth]{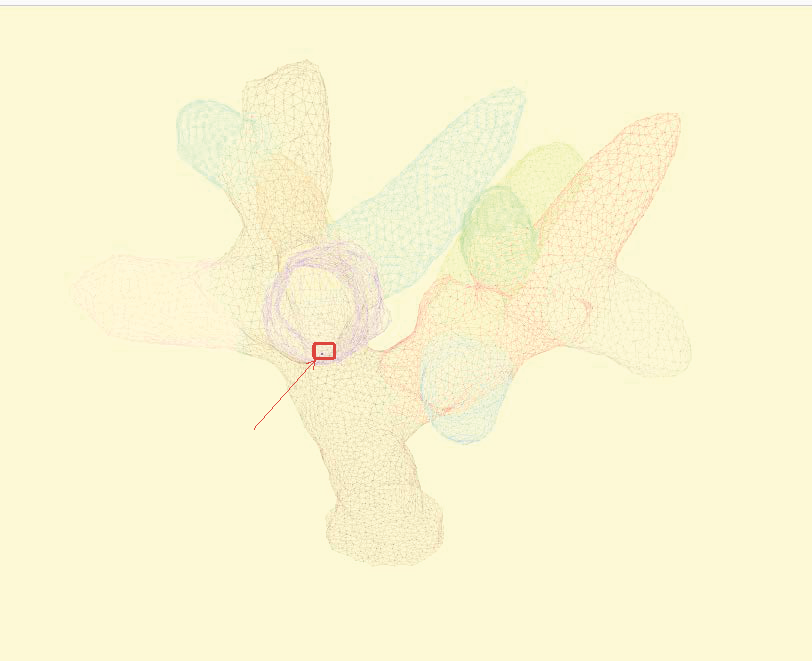}}
\caption{(a) Borders between branches; (b) Example of marked vertex (see arrow) to identify a branch. }\label{fig:coralB}
\end{figure}

The surface mesh coral model is shown in Fig.~\ref{fig:corala} as a visualisation in MeshLab in a reduced resolution with respect to the original Wavefront .obj file. In Fig.~\ref{fig:coralb} the same model is shown after its conversion to the input format for model checking and visualisation with \polyvisualizer{.} The identification of the branches have been obtained through model checking in the following way. First the borders between the branches have been marked using MeshLab (shown in Fig.~\ref{fig:coralBa}). Then some selected vertices of each branch have been marked manually, each with a different mark (shown in Fig.~\ref{fig:coralBb} for one specific branch). 

\begin{figure}
\centering
\fbox{
$
\begin{array}{l c l}
\mbox{\sf b0} &\equiv &  \gamma(\lneg \mbox{\sf border}, \mbox{\sf bSel0})\\
\mbox{\sf b1} &\equiv &  \gamma(\lneg \mbox{\sf border}, \mbox{\sf bSel1})\\
\cdots\\
\mbox{\sf b13} &\equiv &  \gamma(\lneg \mbox{\sf border}, \mbox{\sf bSel13})\\
\end{array}
$
}
\caption{Atomic proposition letters {\sf border}, {\sf bSel0}, {\sf bSel1}, etc. of selected triangular faces for the various branches are assumed given; {\sf b0} through {\sf b13} characterises the cells belonging to the various branches.}\label{fig:SLCScoralA}
\end{figure}

The \slcsG\ formulas to identify the various branches are straightforward and shown in Fig.~\ref{fig:SLCScoralA}. For example, the formula {\sf b1} is satisfied by all cells in the model that are not part of a border and through which one can reach, passing only by such non border elements, a cell satisfying {\sf bSel1}. Providing the visualiser with the model checking results, together with a definition of the colour in which to show the cells satisfying the various properties {\sf b0} to {\sf b13} and the borders, we obtain the polyhedron shown in Fig.~\ref{fig:coralb}. The model checking time for the formulas in Fig.~\ref{fig:SLCScoralA} is 2.7 seconds. The coral model consists of 67,573 vertices, 202,712 edges and 135,141 triangles. The full \imgql\ specifications for the coral example can be found in Appendix~\ref{app:coral}.

Next we are interested in finding the branching hierarchy in the coral structure. Such branching hierarchy provides information about the complexity of the corals. Both the covering of the ocean bottom with corals and particularly its covering with branching corals have found to be positively correlated with the structural complexity of coral reefs. In turn, structural complexity of coral reefs has been shown to positively influence several measures of biodiversity in such reefs. Branching coral cover of the ocean bottom may be particularly likely to contribute fine-scale structural complexity to reefs, which can be important to a range of organisms, such as fish and mobile invertebrates (see e.g.~\cite{GrN2013} page 322). One way to measure the branching hierarchy of a coral is through investigating the parent-child hierarchy of branches. In the following we will use \polylogica\ to colour the various groups of branches according their rank in such a hierarchy. 

The procedure follows two alternating phases. It starts from the root of the coral. We define two properties for each rank level n. The first property, indicated by {\sf rankPn}, defines all the cells that belong to a certain rank level. The root is the first rank level. The second property, indicated by {\sf rankAn}, defines all the cells that belong to rank level n or lower, including the borders in between. We define two derived operators {\sf grow(x,y)} and {\sf throughCom(x,y,z)}.  Informally, the operator {\sf grow(x,y)= x $\lor$ (y $\land \gamma$(y,x))} extends an area of cells each satisfying $x$ with an adjacent area of cells each satisfying $y$. The operator {\sf throughCom(x, y, z)} combines the $\gamma$ operator with the {\sf grow} operator. Informally, this formula is satisfied by any cell from which a \plm-path starts that passes by cells satisfying $x$ and reaches a cell satisfying {\sf grow(z,y)}. When instantiated as {\sf throughCom($\neg$border, border, rank)}, this operator is used to find cells in the branches, that are not border cells but via which one can reach the border to cells in branches with a lower rank. We also define an auxiliary operator {\sf difference(x,y)}. The latter property is satisfied by cells that satisfy {\sf x} but not {\sf y}. The specification in Fig.~\ref{fig:SLCScoralB} shows the formulas for identifying three rank levels. Of course, these can be extended in the obvious way to identify rank of branches in coral with a larger branching structure. The specification assumes that the atomic proposition letters {\sf border} and {\sf rootSel}, some selected cells of the root branch, are given. Both were obtained by annotating the model with the tool MeshLab. 

\begin{figure}
\centering
\fbox{
$
\begin{array}{l c l}
\mbox{\sf grow(x,y)} &\equiv & \mbox{\sf x} \lor (\mbox{\sf y} \land \gamma(\mbox{\sf y},\mbox{\sf x}))\\
\mbox{\sf throughCom(x,y,z)} &\equiv & \gamma(\mbox{\sf x}, \mbox{\sf grow}(\mbox{\sf z},\mbox{\sf y}))\\
\mbox{\sf difference(x,y)} &\equiv & \mbox{\sf x} \land \neg\mbox{\sf y}\\
\mbox{\sf root}  &\equiv & \gamma(\neg\mbox{\sf border}, \mbox{\sf  rootSel})\\
\mbox{\sf rankP1} &\equiv &  \mbox{\sf root}\\ %let rankP1 = root
\mbox{\sf rankA1} &\equiv &  \mbox{\sf root}\\
\mbox{\sf rankP2} &\equiv &  \dirnear(\mbox{\sf difference}(\mbox{\sf throughCom}(\neg \mbox{\sf border}, \mbox{\sf border}, \mbox{\sf rankP1}), \mbox{\sf rankA1}))\\
\mbox{\sf rankA2} &\equiv & \mbox{\sf rankP2} \lor \mbox{\sf grow}(\mbox{\sf rankA1},\mbox{\sf border})\\
\mbox{\sf rankP3} &\equiv &   \dirnear(\mbox{\sf difference}(\mbox{\sf throughCom}(\neg \mbox{\sf border}, \mbox{\sf border}, \mbox{\sf rankP2}), \mbox{\sf rankA2}))\\
\mbox{\sf rankA3} &\equiv & \mbox{\sf rankP3} \lor \mbox{\sf grow}(\mbox{\sf rankA2},\mbox{\sf border})
\end{array}
$
}
\caption{Finding the rank of coral branches. Atomic proposition letters {\sf border} and {\sf rootSel} are assumed given.}\label{fig:SLCScoralB}
\end{figure}

\begin{figure}[h!]
\centering
%\subfloat[]{\includegraphics[width=0.38\textwidth]{figs/coral_ranks.png}}\quad\quad\quad
\subfloat[\label{fig:coralDa}]{\includegraphics[width=0.38\textwidth]{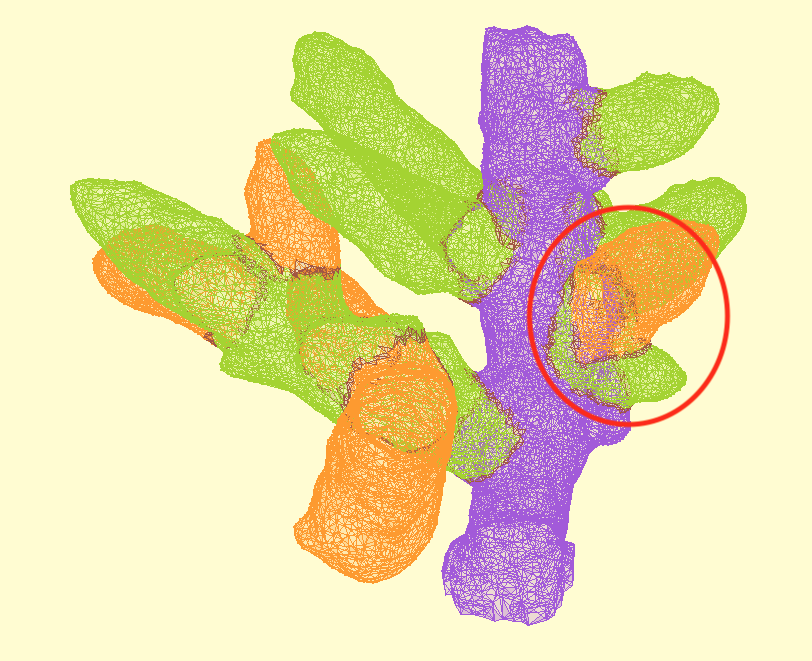}}\quad\quad\quad
\subfloat[\label{fig:coralDb}]{\includegraphics[width=0.38\textwidth]{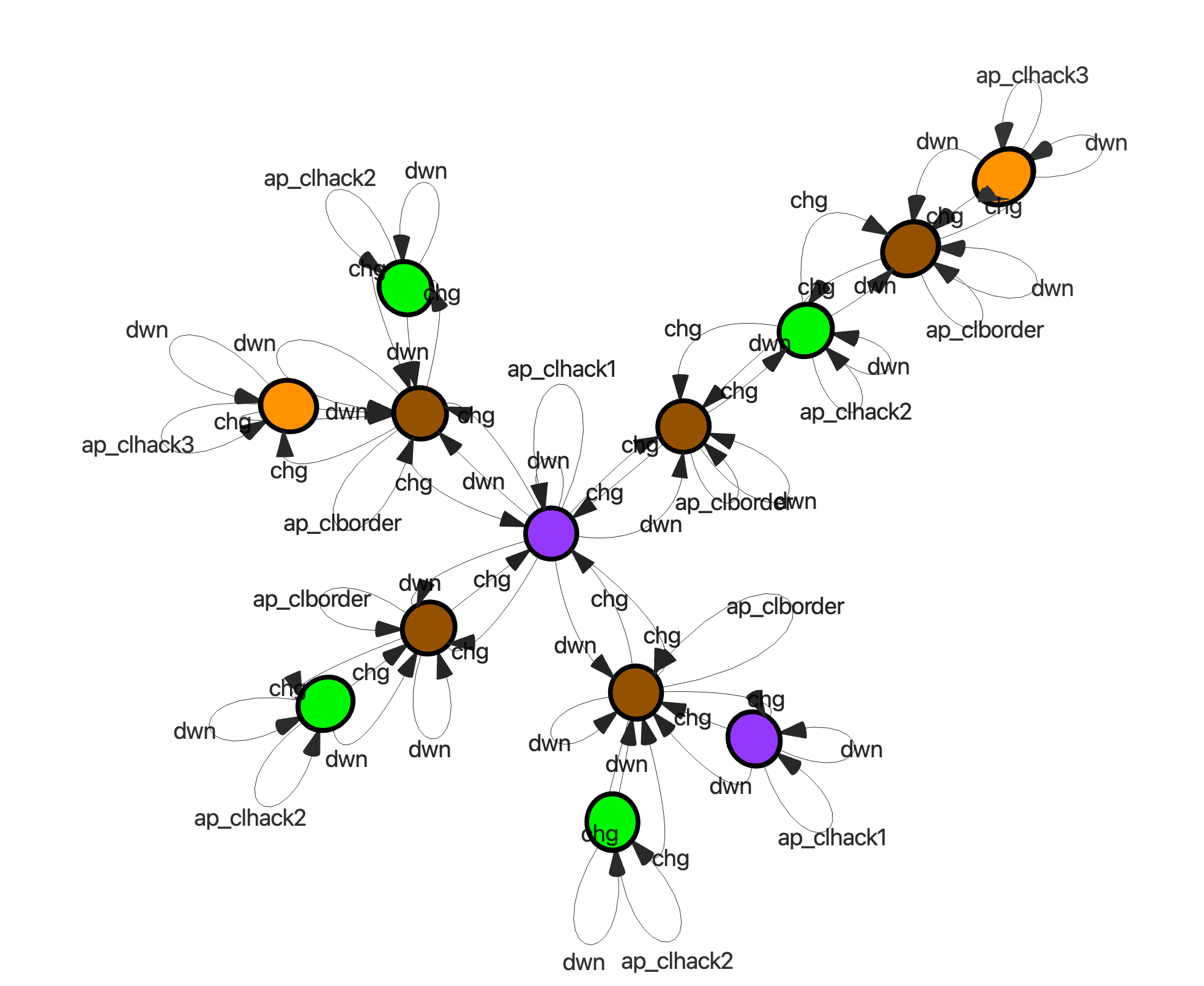}}
\caption{(a) Rank levels in the same coral (turned 180 degrees around its vertical axes w.r.t. Fig.~\ref{fig:coral}); (b) Minimised model as LTS.  Red circle shows two branches sharing part of their border.}\label{fig:coralD}
\end{figure}

The model checking results of the specification in Fig.~\ref{fig:SLCScoralB} is shown in Fig.~\ref{fig:coralDa}. Rank level {\sf rankP1} is shown in purple, rank level {\sf rankP2} in green and rank level {\sf rankP3} in orange. Borders between branches are shown in brown. The model checking time for the formulas in Fig.~\ref{fig:SLCScoralB} is 6 seconds, loading the file takes 16.5 seconds. 
%The coral model consists of 67,573 vertices, 202,712 edges and 135,141 triangles. 
The full \imgql\ specifications for the coral example can be found in Appendix~\ref{app:coral}.

One might be interested also in the core structure of a branching coral. There are several known computer graphics methods that can be used to obtain such a structure, for instance methods that extract a topological `skeleton' from a triangular surface mesh, i.e. a sort of thin version of the shape that is equidistant to its borders. In our setting we could obtain an abstract version of the structure by minimising the structure w.r.t. $\wplmbis$, the weaker bisimulation related to \slcsE\ (see Section~\ref{sec:BackAndNotat}). The result of minimising the structure in Fig.~\ref{fig:coralDa} is shown as a Labelled Transition System (LTS) in Fig.~\ref{fig:coralDb}, where the states in the LTS have been given colours that correspond to the rank levels in Fig.~\ref{fig:coralD}a. The labels of the self-loops in the LTS, such as {\sf ap\_b1} and {\sf ap\_border}, denote the atomic propositions of the states, the other transition labels, {\sf chg} and {\sf dwn}, are auxiliary labels used in the minimisation procedure (see~\cite{Be+24b} for details).  This LTS does not give exactly the same information as a topological skeleton, but may reveal other interesting aspects. In the specific case of the rank levels of this coral, for example, it reveals that a second rank (green) shares part of its border (brown) with a third rank (orange) and the root rank (purple), as shown in the left-upper part of Fig.~\ref{fig:coralDb}. This situation is indeed present in the coral and can be seen in Fig.~\ref{fig:coralDa}, indicated by the red circle. This image of the coral is the same as in Fig.~\ref{fig:coral} but turned 180 degrees around its vertical axes. This may indicate an ambiguity in how to count the number of rank levels as this requires to answer the question which branch is growing from which other branch.

But let us first illustrate how we obtain the minimal LTS model of a coral. First of all, we use \polylogica\ to find the rank levels as described previously. The model checking result for each rank level can be saved in a simple .json format that provides for each cell in the model true, in case it is part of that rank level, and false if it is not. We can use this outcome to add new atomic propositions to the original model. For example, in case we have the results for rank level 1, we add the atomic proposition {\sf "rank1"} to each cell in the model file, replacing the atomic propositions that were already there. This is easy to do in an automatic way using a simple python script because in the result file of model checking the results are given in exactly the same order as the cells in the model file. If we do this for every rank, we get a model with only {\sf rank1}, {\sf rank2}, {\sf rank3} and {\sf border} as atomic propositions, and each cell is labelled by exactly one atomic proposition. Applying the toolchain described in~\cite{Be+24b} for minimising the model modulo $\wplmbis$, the weaker bisimulation related to \slcsE{,} we obtain the minimised LTS as shown in Fig.~\ref{fig:coralDb} (apart from colouring the states, which was done manually to facilitate the interpretation of the state labels).

\begin{figure}[t!]
\centering
%\subfloat[]{\includegraphics[width=0.38\textwidth]{figs/coral_ranks.png}}\quad\quad\quad
\subfloat[\label{fig:coralEa}]{\includegraphics[width=0.38\textwidth]{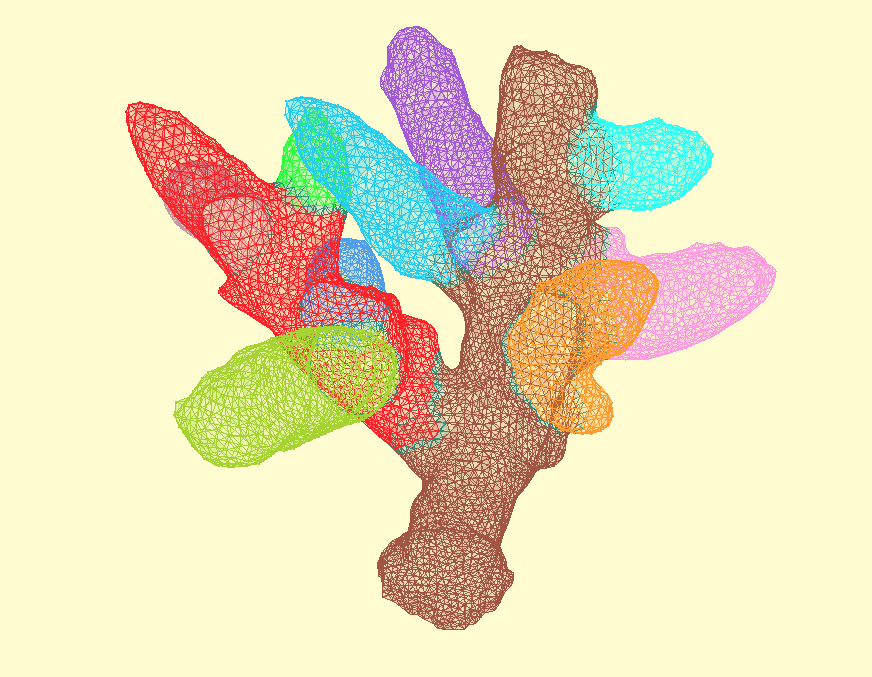}}
\subfloat[\label{fig:coralEb}]{\includegraphics[width=0.68\textwidth]{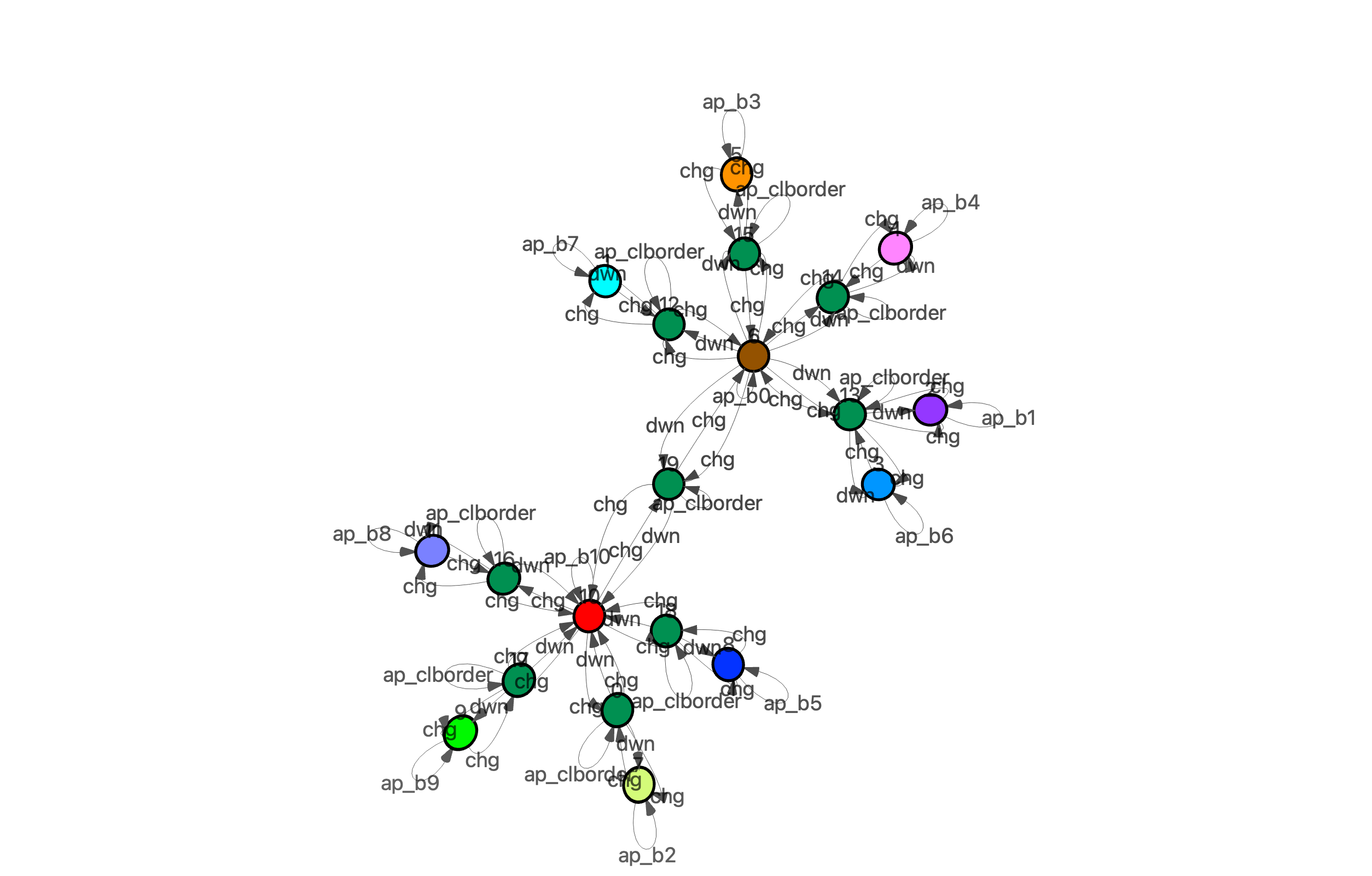}}
\caption{(a) Branches of the coral with individual colours; (b) Minimised model as LTS with state colours reflecting that of corresponding branches in (a), which states representing borders in dark green.}\label{fig:coralE}
\end{figure}

We can perform a similar analysis on the coral model in which all branches were given a different colour as in Fig.~\ref{fig:coralb}.  The result of this minimisation is the LTS shown in Fig.~\ref{fig:coralEb}. In the latter the colours of the states representing equivalence classes reflect the colours of the branches of the coral in Fig.~\ref{fig:coralEa}. The states representing the border classes are in dark green. This makes it easy to recognise the full structure of the coral reflected in the minimised LTS model. A closer look also reveals an interesting aspect. In the upper right part of the LTS the blue branch and the purple branch are shown to share a border. This was not the case in the minimal model in Fig.~\ref{fig:coralDb}, since there these two branches had the same colour because it belonged to the same rank. Furthermore, the orange branch is now a single branch. The ranking procedure had split that branch into two different parts, assigning different ranks.

Note that the above analyses are presenting an alternative use of model minimisation. Usually one is using minimisation to increase the efficiency of spatial model checking for large models~\cite{Be+24}. Actually, the possibility to add (or replace) atomic propositions in a model as we did above can be of great use also in other setting, as it makes it possible to insert intermediate model checking results in the model itself so that further formulas to check can be of much shorter length, saving precious computing resources.

\subsection{Studying Curvature Properties in a Simple Aircraft Model}

% See for further info on use of MeshLab also the following link:
% https://stackoverflow.com/questions/59194552/meshlab-does-the-topology-of-my-mesh-effect-the-curvature-results/59266954

\begin{figure}[b!]
\centering
\subfloat[\label{fig:ufoa}]{\includegraphics[width=0.45\textwidth]{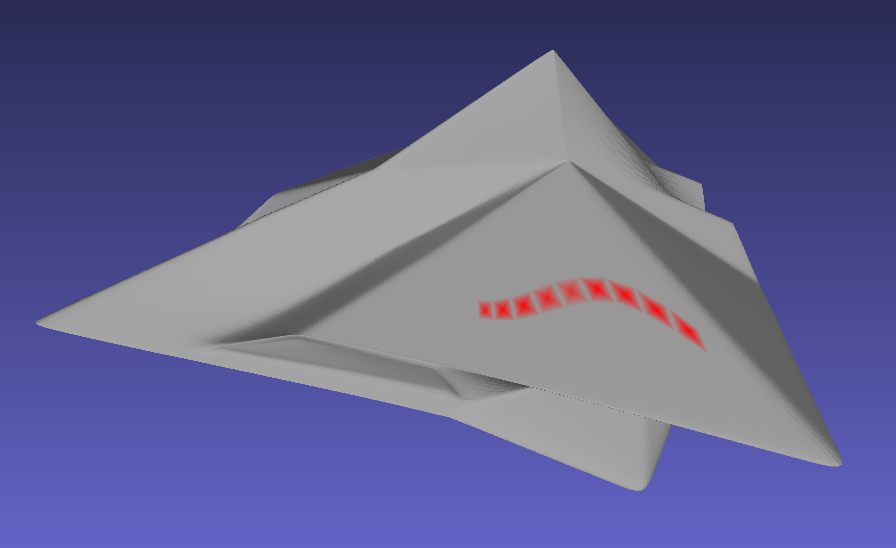}}\quad\quad\quad
\subfloat[\label{fig:ufob}]{\includegraphics[width=0.45\textwidth]{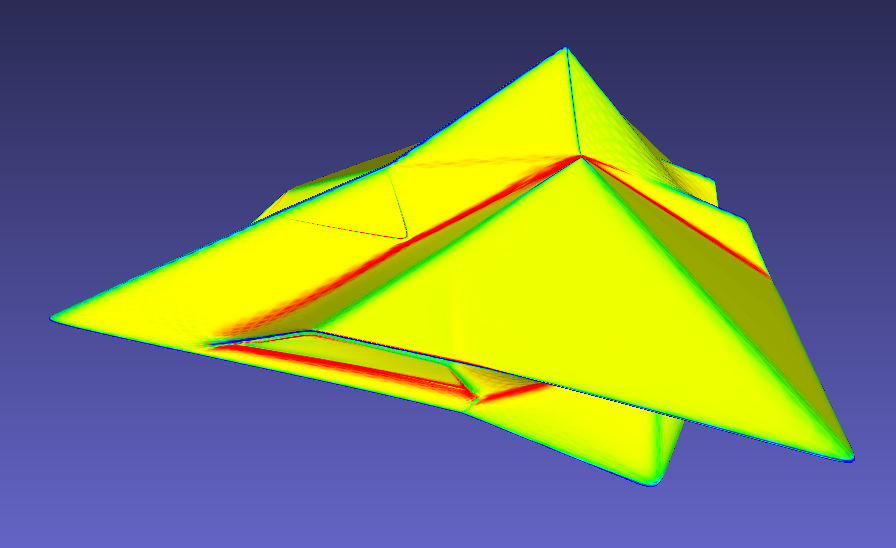}}\\
\subfloat[\label{fig:ufoc}]{\includegraphics[width=0.45\textwidth]{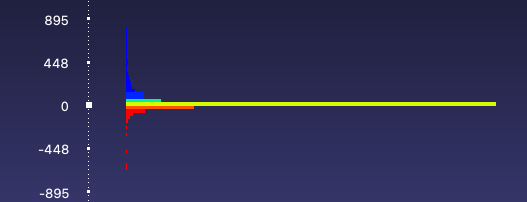}}
\caption{(a) Aircraft in MeshLab~\cite{MeshLab2008}; (b) Curvatures of the toy aircraft as computed by MeshLab~\cite{MeshLab2008}: Flat surface in yellow, concave edges in red, convex edges in green and sharp convex edges in blue, as also shown in the histogram of curvature values for vertices (c).}\label{fig:ufo}
\end{figure}

For the third example we use a simple aircraft model to illustrate how PolyLogicA can be used to analyse other aspects than plain cell colours of an object. In particular, we can convert other information of the model into colours and then analyse the result with PolyLogicA. The aircraft is composed of simple surfaces that are connected together. Such connections form angles that can be concave or convex and that can be more or less sharp. The surfaces themselves are mainly flat, but there are some small ridges as well. A view of the toy aircraft is shown in Fig.~\ref{fig:ufoa} and a version in which concave and convex edges are shown in different colours is presented in Fig.~\ref{fig:ufob}. 

\begin{figure}[t!]
\centering
\subfloat[\label{fig:polyvisa}]{\includegraphics[width=0.45\textwidth]{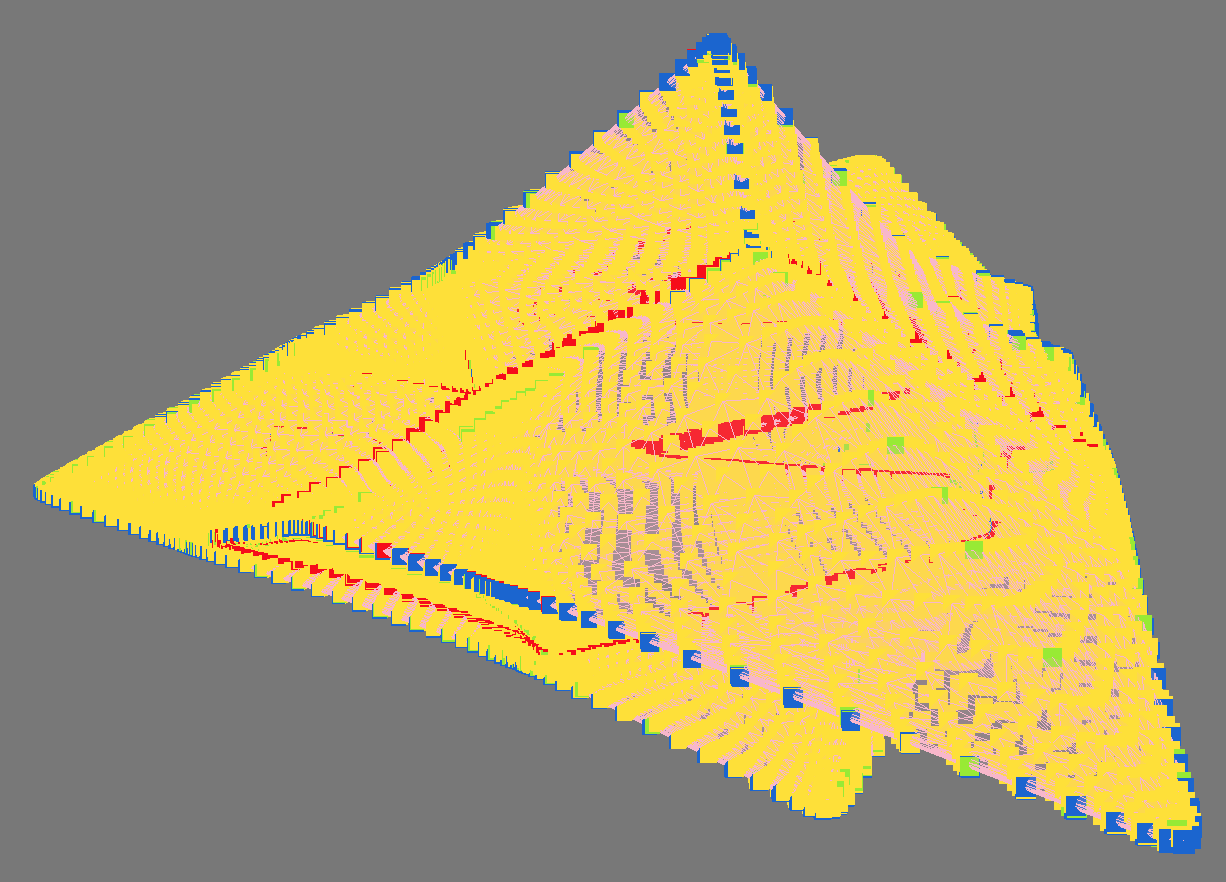}}\quad\quad\quad
\subfloat[\label{fig:polyvisb}]{\includegraphics[width=0.45\textwidth]{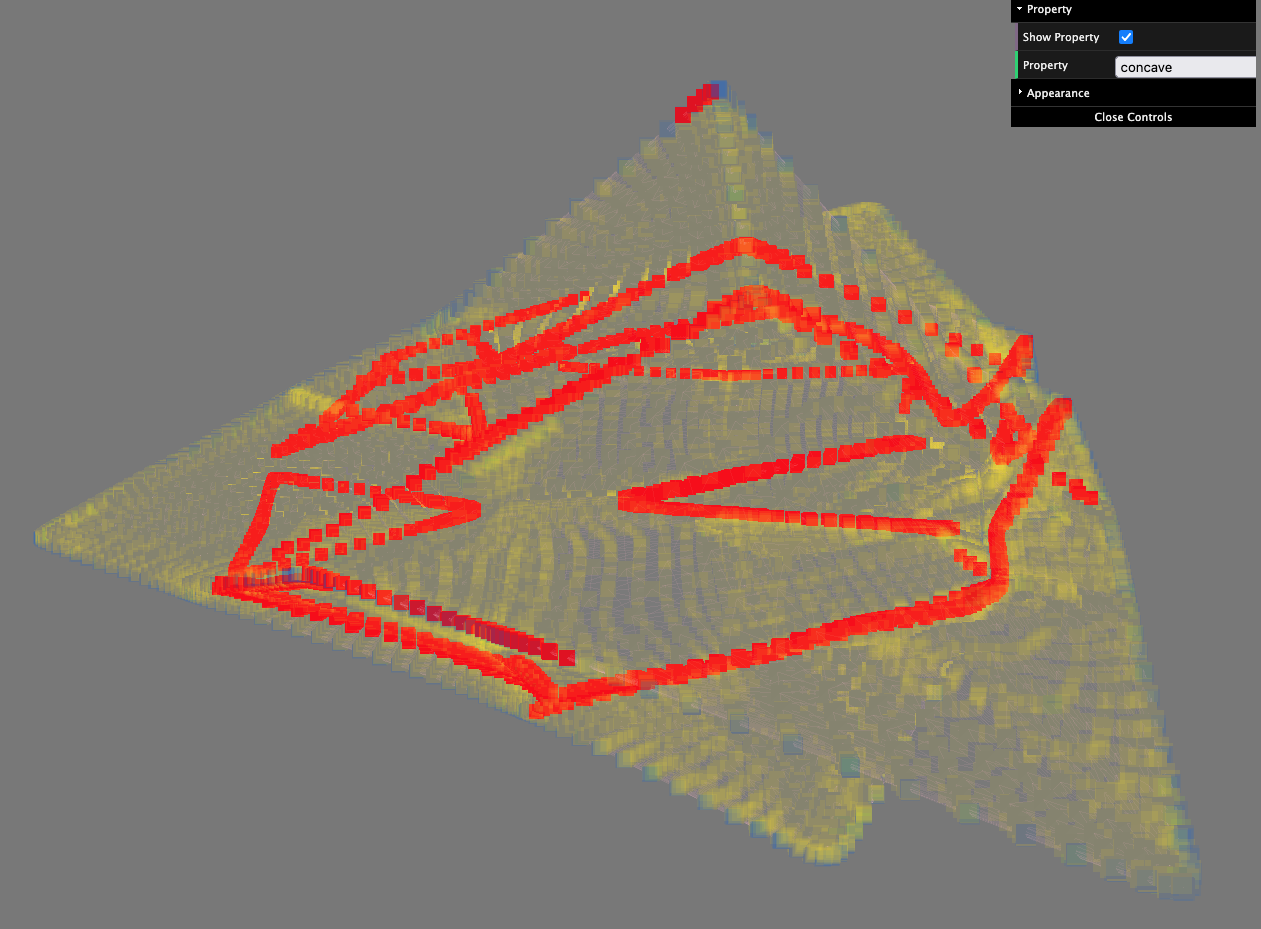}}
\caption{(a) Aircraft in \polyvisualizer{;} (b) Vertices satisfying `concave' highlighted in red. }\label{fig:polyvis}
\end{figure}

\begin{figure}[t!]
\centering
\subfloat[\label{fig:polyvisBa}]{\includegraphics[width=0.45\textwidth]{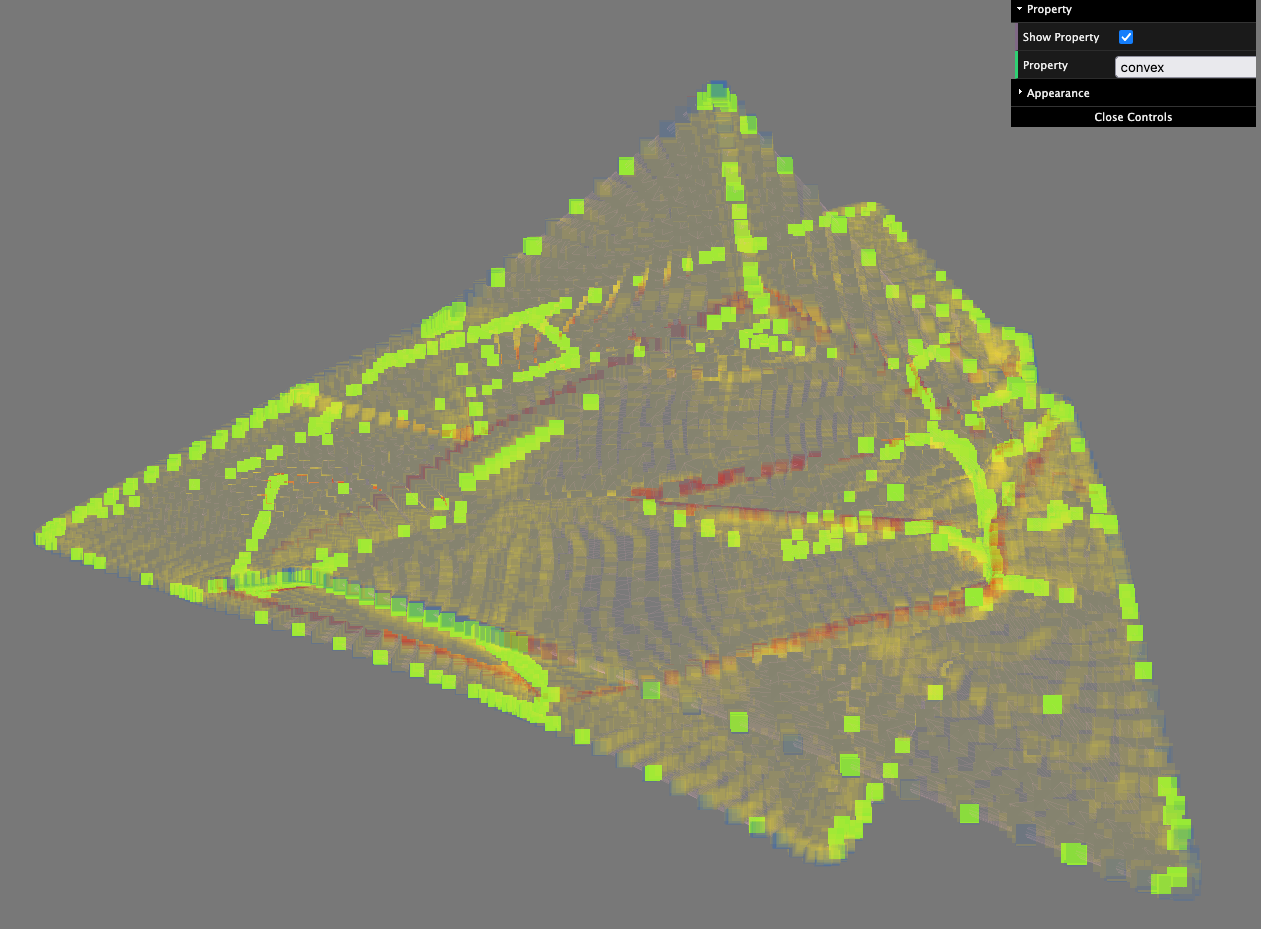}}\quad\quad\quad
\subfloat[\label{fig:polyvisBb}]{\includegraphics[width=0.45\textwidth]{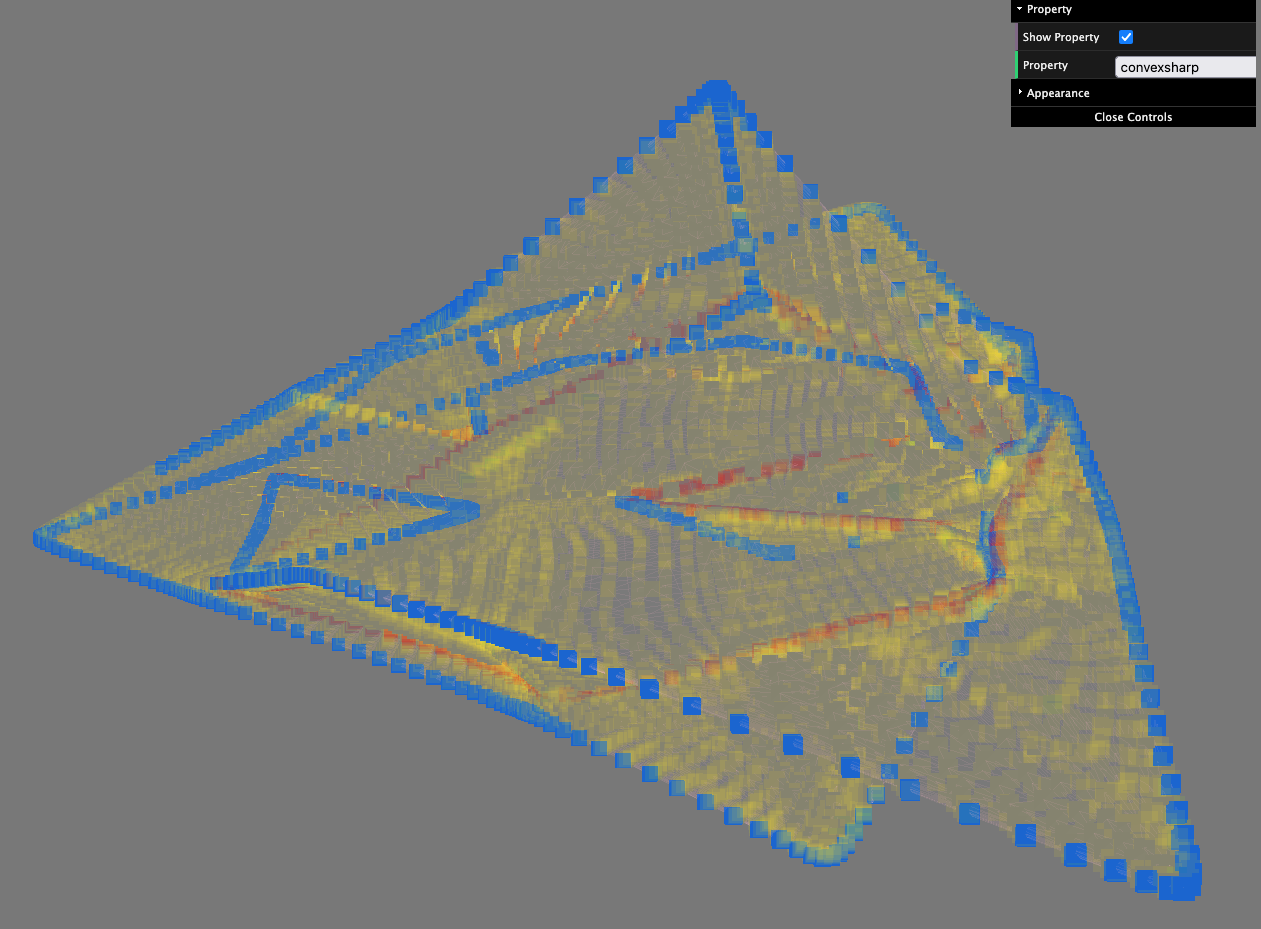}}
\caption{(a) Vertices satisfying `convex' highlighted in green; (b) Vertices satisfying `convexsharp' highlighted in blue. }\label{fig:polyvisB}
\end{figure}

The image in Fig.~\ref{fig:ufob} has been obtained by computing discrete curvatures with MeshLab~\cite{MeshLab2008}. The resulting image can be saved as a Wavefront .obj image in which the values of the curvature are saved as colours for each {\em vertex}. This .obj image, in turn, can be converted, using a dedicated python script, into an input suitable for polyhedra model checking with \polylogica\ and for viewing with the \polyvisualizer{.}  In the python script we can define the relation between the colours of the vertices and the names of the atomic propositions. For example, we defined four atomic propositions, with the names `{\sf flat}', `{\sf concave}', `{\sf convex}' and `{\sf convexsharp}', corresponding to the four colours used in the histogram in Fig.~\ref{fig:ufo}c that was also produced with MeshLab by rendering the curvature measure as a vertex quality.  A vertex (or point) gets the atomic proposition `{\sf concave}' if it is red (in rgb terms [255, 0, 0]) with a margin of 30. Similarly for {\sf convex} (rgb [0, 255, 0]) and {\sf convexsharp} (rgb [0, 0, 255]). The latter represents a high curvature value indicating very sharp convex curves. We gave all remaining vertices the atomic proposition `{\sf flat}' and all remaining triangles and segments of the aircraft the atomic proposition `{\sf ufo}'. The result of this conversion is shown in the visualisation of the .json model in Fig.~\ref{fig:polyvis}a, where we chose to show vertices satisfying {\sf flat} in yellow, vertices satisfying {\sf concave} in red, and so on.
We can now use \polylogica\ on the generated .json model to visualise the atomic propositions one by one. For example, in Fig.~\ref{fig:polyvisb} all vertices satisfying {\sf concave} are highlighted in red. In Fig.~\ref{fig:polyvisB} vertices satisfying {\sf convex} and {\sf convexsharp} are shown, highlighted in green (Fig.~\ref{fig:polyvisBa}) ad highlighted in blue (Fig.~\ref{fig:polyvisBb}), respectively.

\begin{figure}
\centering
\fbox{
$
\begin{array}{l c l}
%\dirnear(x) &\equiv & \gamma(x, \ltrue)\\
%\mbox{\sf reach(x,y)} &\equiv & x \land \gamma(x,y) \\
%\mbox{\sf sur(x,y)} &\equiv & x \land \neg(\gamma(\lneg(y),\lneg(x \lor y))))\\
\mbox{\sf grow(x,y)} &\equiv &  x \lor (y \land \gamma(y, x))\\
%\mbox{\sf moreconcave} &\equiv &  \cnvnear(\mbox{\sf concave})\\
%\mbox{\sf moreflat} &\equiv &  \cnvnear(\mbox{\sf flat})\\
%\mbox{\sf moreconvexsharp} &\equiv & \cnvnear({\sf convexsharp})\\
\mbox{\sf convexmeetsconcave} &\equiv & \mbox{\sf convex} \land \gamma(\cnvnear\mbox{\sf concave},\mbox{\sf convex})\\
\mbox{\sf ridgeborder} &\equiv & \mbox{\sf concave} \land \gamma(\cnvnear{\sf convexsharp}, \mbox{\sf concave})\\
%\mbox{\sf moreridgeborder} &\equiv & \cnvnear({\sf ridgeborder})\\
%\mbox{\sf ridgearea} &\equiv & \mbox{\sf ufo} \land \mbox{\sf sur}((\mbox{\sf moreconvex} \lor \mbox{\sf moreconvexsharp} \lor \mbox{\sf ufo}),\mbox{\sf moreridgeborder})\\
\mbox{\sf flatridgearea} &\equiv & \cnvnear\mbox{\sf flat} \land (\lneg \mbox{\sf grow}((\mbox{\sf ridgeborder} \land \mbox{\sf concave}), \cnvnear\mbox{\sf flat})).
\end{array}
$
}
\caption{Derived \slcsG\ operator {\sf grow(x,y)};  \slcsG\ formulas expressing the properties  {\sf convexmeetsconcave}, {\sf ridgeborder} and {\sf flatridgearea}; atomic proposition letters $\mbox{\sf concave},\mbox{\sf convex},\mbox{\sf convexsharp}$ and $\mbox{\sf flat}$ are assumed given.\label{fig:SLCSufo}}
\end{figure}

\begin{figure}[t!]
\centering
\subfloat[\label{fig:polyvisCa}]{\includegraphics[width=0.45\textwidth]{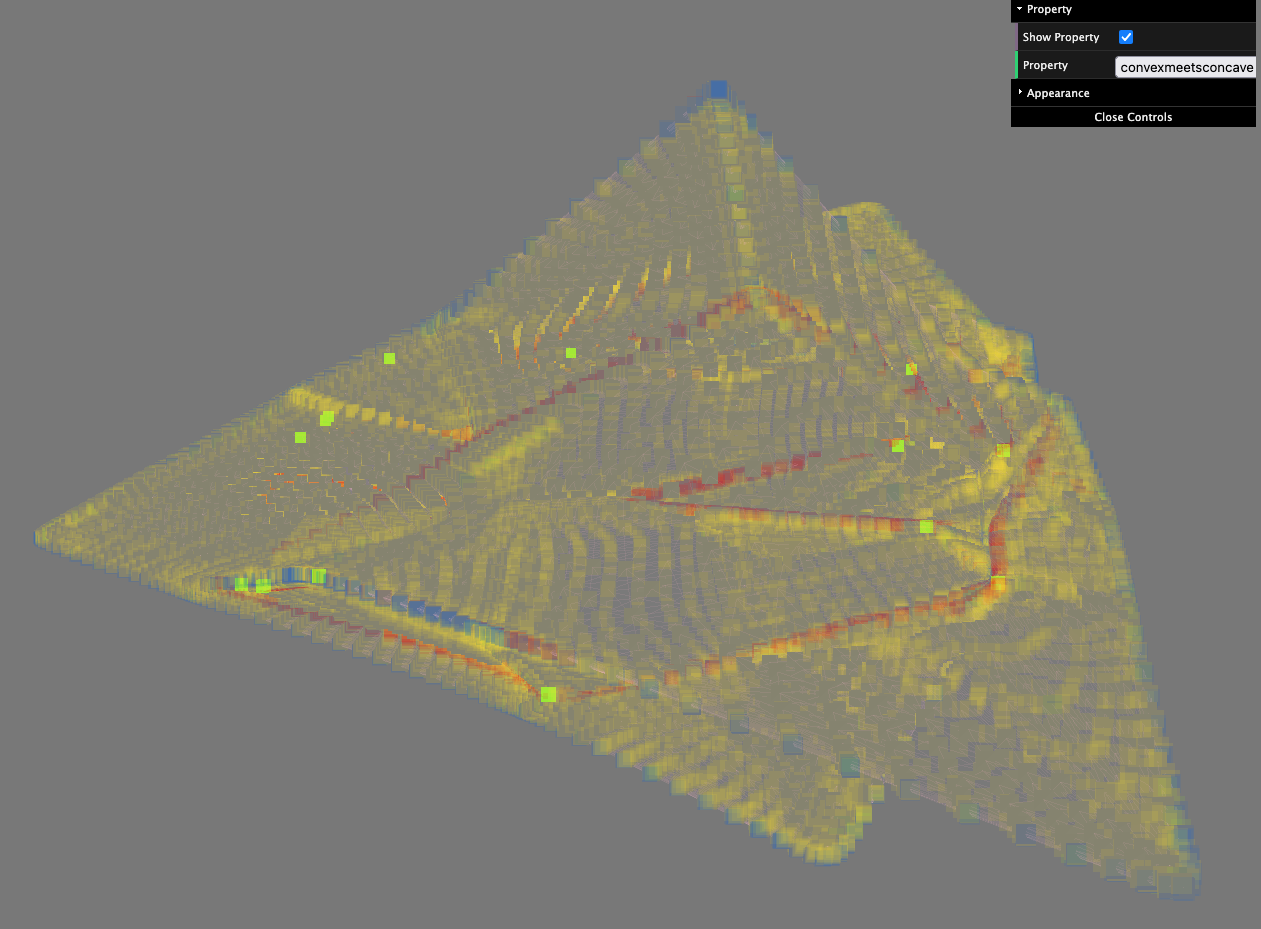}}\quad\quad\quad
\subfloat[\label{fig:polyvisCb}]{\includegraphics[width=0.45\textwidth]{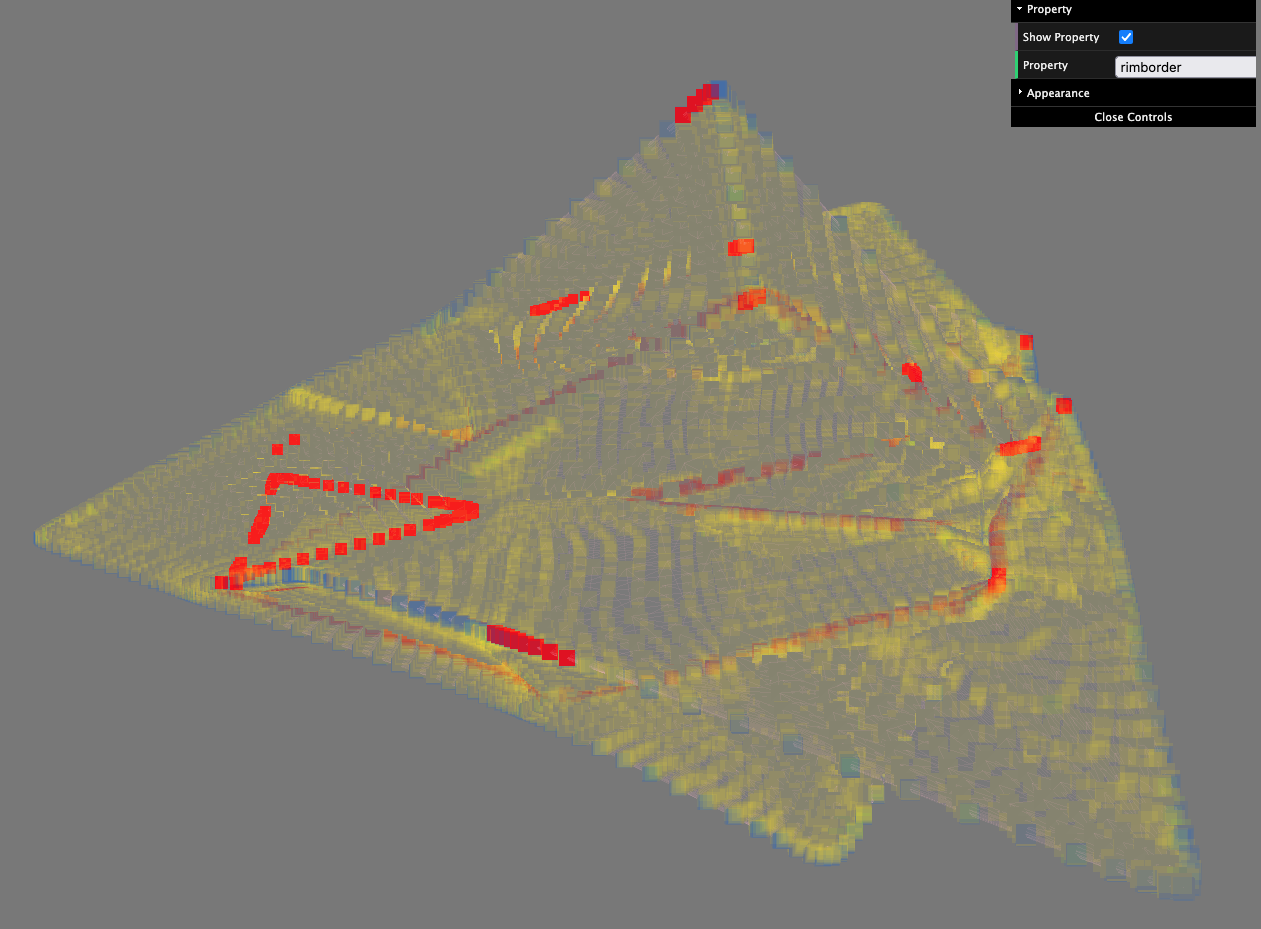}}
\caption{(a) {\sf convex} meeting {\sf concave} highlighted in green; (b) Vertices satisfying {\sf ridgeborder} highlighted in red. }\label{fig:polyvisC}
\end{figure}

More interesting properties are specified in Fig.~\ref{fig:SLCSufo}. For example, we might be interested in convex vertices that are close to concave ones. 
The formula {\sf $\cnvnear$concave} is satisfied by all edges and triangles that are connected to a vertex satisfying `concave'. We can then define {\sf convexmeetsconcave = convex $\land$ $\gamma$($\cnvnear$concave,convex)}. The model checking result for this formula is shown in Fig.~\ref{fig:polyvisCa} by the highlighted green vertices. 
 
\begin{figure}[t!]
\centering
\subfloat[\label{fig:polyvisDa}]{\includegraphics[width=0.38\textwidth]{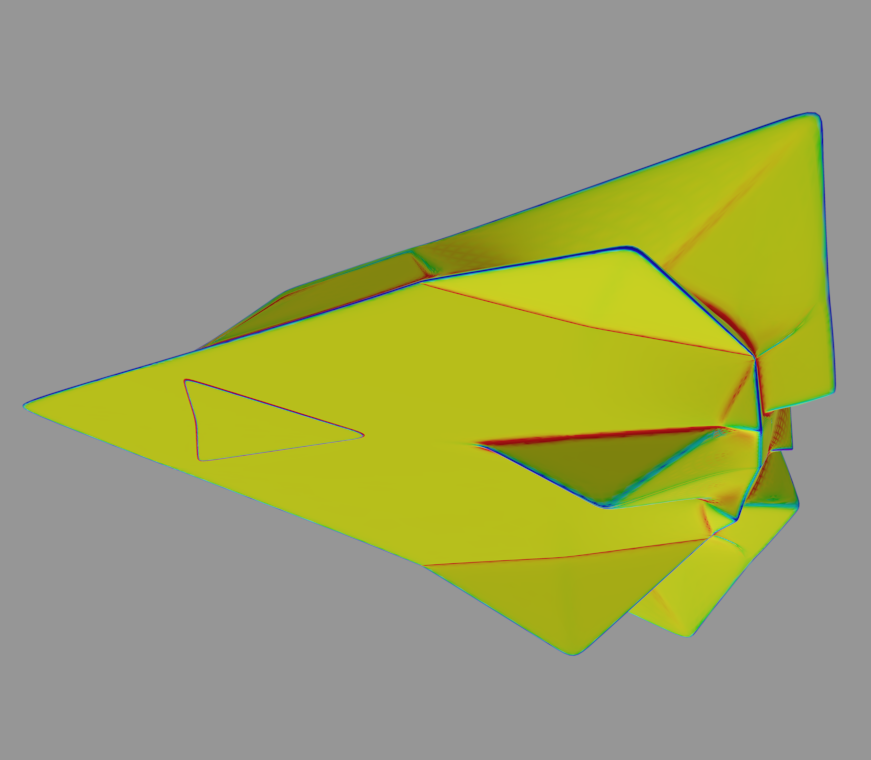}}\quad\quad\quad
\subfloat[\label{fig:polyvisDb}]{\includegraphics[width=0.45\textwidth]{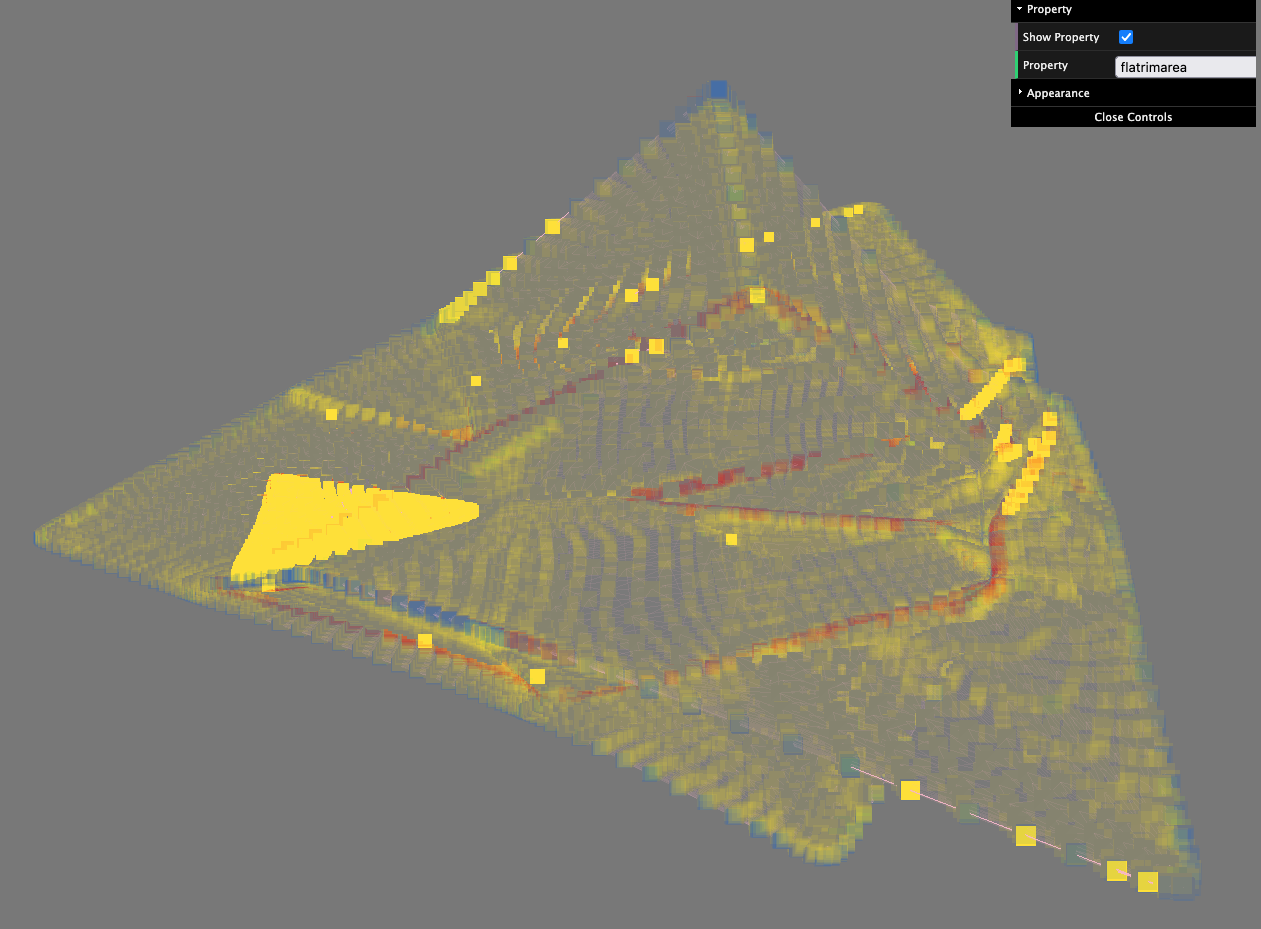}}
\caption{(a) Bottom view of aircraft; (b) Vertices satisfying `{\sf flatridgearea}' highlighted in yellow. }\label{fig:polyvisD}
\end{figure}

A rather subtle property is to find areas that are surrounded by a small, sharp ridge. We see such an area for example as a triangle shape in the bottom of the aircraft as shown in Fig.~\ref{fig:polyvisDa}. Such ridges are characterised by a thin concave area next to a thin sharp convex area. One can define such a ridge as {\sf ridgeborder = concave} $\land\; \gamma( \cnvnear{\sf convexsharp}, \mbox{\sf concave})$ and then look for an area surrounded by such a ridge defining {\sf flatridgearea} = $ \cnvnear\mbox{\sf flat} \land\; (\neg $ {\sf grow((ridgeborder} $\land$ {\sf concave)}, $\cnvnear\mbox{\sf flat}))$, where $\cnvnear\mbox{\sf flat}$ are the vertices satisfying {\sf flat} extended with their directly adjacent edges and triangles. The result of \polylogica\ model checking of this formula is shown in Fig.~\ref{fig:polyvisDb}, where the area that satisfies the property is highlighted in yellow. It is easy to see that besides the triangular shape in the bottom of the aircraft there are also other small areas that have this property in the rear of the aircraft and along the edges of the wings. This illustrates how \polylogica\ could be used to find also otherwise not easy to detect aspects of the curvature of an object, which might be very useful, for example, when looking for small defects in constructions.

Just to provide an indication of the performance, the total analysis time of all the properties in Fig.~\ref{fig:SLCSufo} together amounts to about 10 seconds, including loading of the input file and writing of the results. The pure model checking time takes 3 seconds. The model consists of 31,810 vertices and a total of 63,615 cells. The full \imgql\ specifications for the aircraft can be found in Appendix~\ref{app:ufo}.

\section{Conclusion}\label{sec:conclusion}

We have shown various ways in which polyhedral model checking can be used to analyse aspects of triangular surface meshes and volume meshes that can be found in the domain of computer graphics. To this purpose, we have used our spatial model checker \polylogica\ in combination with the well-established computer graphics tool MeshLab~\cite{MeshLab2008}. The latter provides a huge library of operations on surface and volume meshes, among which manual annotation of objects and operations such as computing the curvature characteristics of objects and many more. The results of such operations can again be saved as a mesh object and then converted into the input format of the model checker inserting useful atomic propositions in the model. This way models can be analysed from a wide range of perspectives. This paper has only illustrated a small subset of the possibilities. 

In future work we plan to address further models from various domains, illustrating also how \polylogica\ could be used to check properties of more than one perspective (e.g. curvature, distance, texture) of a model at once. Recent work on geometric analysis through spatial logics to check the consistency of 3D geological models~\cite{PKBH2025} also seems a promising domain of application, besides a continuation of the application of polyhedral model checking in the medical domain~\cite{Be+22}.

\begin{credits}
\subsubsection{\ackname}
Research partially supported by
Bilateral project between CNR (Italy) and SRNSFG (Georgia) ``Model Checking for Polyhedral Logic'' (\#CNR-22-010);
European Union - Next GenerationEU - National Recovery and Resilience Plan (NRRP), Investment 1.5 Ecosystems of Innovation, Project “Tuscany Health Ecosystem” (THE), CUP: B83C22003930001;
European Union - Next-GenerationEU - National Recovery and Resilience Plan (NRRP) – MISSION 4 COMPONENT 2, INVESTMENT N. 1.1, CALL PRIN 2022 D.D. 104 02-02-2022 – (Stendhal) CUP N. B53D23012850006; 
MUR project
PRIN 2020TL3X8X ``T-LADIES''; CNR project "Formal Methods in Software Engineering 2.0", CUP B53C24000720005.
%The authors are listed in alphabetical order, as they equally contributed to the work presented in this paper.

\subsubsection{\discintname}
The authors have no competing interests to declare that are
relevant to the content of this article.

\end{credits}
\bibliographystyle{splncs04}
\bibliography{ARXIV_POLY_EXAMPLES_2025.bib}
%
%\pagebreak
%\appendix

\section{Appendix}\label{sec:app}

\subsection{Full \imgql\ Specification for the 3D Maze Example}\label{app:maze}

{\small
\begin{verbatim}
load model = "polylogica/3DMAZE/3DmazeModel.json"

let sur(x,y)	= x & (!(through(!y,!(x | y))))

let green       = ap("G")
let white       = ap("W")
let black       = ap("B")
let red         = ap("R")
let corridor    = ap("corridor")

let blackOrWhite        = (black | white)

let corridorWW			= through( corridor, white ) &
                     !through( corridor, green | black | red )
let corridorWG			= through( corridor, white ) & through( corridor, green )
let corridorWR			= through( corridor, white ) & through( corridor, red )
let corridorWB			= through( corridor, white ) & through( corridor, black )

let whiteToGreen		= through((white | corridorWW | corridorWG), green)
let connWG			= whiteToGreen | through(green,whiteToGreen)
let connRWG			= through((connWG | corridorWR), red) | 
                     through((red | corridorWR), connWG)

// The following two expressions are both identifying white areas surrounded
// completely by corridors leading to black rooms
let whiteNoGreen		= (white | corridorWW) & !whiteToGreen
let whiteSblack			= sur((white | corridorWW), corridorWB) 

save "green" green
save "white" white
save "black" black
save "red" red
save "corridor" corridor
save "corridorWW" corridorWW
save "corridorWG" corridorWG
save "corridorWR" corridorWR
save "corridorWB" corridorWB
save "whiteSblack" whiteSblack
save "whiteNoGreen" whiteNoGreen
save "whiteToGreen" whiteToGreen
save "connWG" connWG
save "connRWG" connRWG
\end{verbatim}
}

\subsection{Full \imgql\ specification for the Coral Example}\label{app:coral}

Specification in \imgql\ for finding the branches in a tree coral.

{\small
\begin{verbatim}
// standard library ---------------------------------------------------------------------

// all cells which have a certain property x or than can reach property y
// through a path of cells that satisfy x
let reach(x,y) = or(x, through(x,y))

// Define eta in terms of gamma (through):
let eta(x,y) = x & through(x,y)

// Define classical closure in terms of gamma (through):
let closure(x) = through(x, tt)

// all the cells which have a certain property x or have a property y and can
// reach cells with property x only through cells of property y
let grow(x,y) = or(x, and(y, through(y, x)))

// all cells which have either property x or property y but not both
//let xor(x,y) = or( and(x, not(y)), and(not(x), y) )

// set operators
let complement(x) = not(x)
let union(x,y) = or(x,y)
let intersection(x,y) = and(x,y)
let difference(x,y) = and(x, not(y))
let symmetricDifference(x,y) = xor(x,y)

//code start
//load model = "./polylogica/EDR26_TEST/EDR26_T6_segmented_low3.json"
load model = "./polylogica/CORAL_Mke/EDR26_T6_segmented_low3.json"


// face properties ----------------------------------------------------------------------
// we expand them to the adjacent vertices and edges using the closure operator
let clborder = closure(ap("border"))


// vertex properties --------------------------------------------------------------------
// here a subset of vertices of each branch has been manually marked and it is used to
// identify the set of cells belonging to each branch.
let rootSel = ap("root")
let bSel0 = ap("b0")
let bSel1 = ap("b1")
let bSel2 = ap("b2")
let bSel3 = ap("b3")
let bSel4 = ap("b4")
let bSel5 = ap("b5")
let bSel6 = ap("b6")
let bSel7 = ap("b7")
let bSel8 = ap("b8")
let bSel9 = ap("b9")
let bSel10 = ap("b10")
let bSel11 = ap("b11")

// compute the cells that belong to each branch -------------------------------------
// these are the cells which reach the manually tagged faces for each branch without
// crossing borders
let root = through(not(clborder), rootSel)
let b0 = through(not(clborder), bSel0)
let b1 = through(not(clborder), bSel1)
let b2 = through(not(clborder), bSel2)
let b3 = through(not(clborder), bSel3)
let b4 = through(not(clborder), bSel4)
let b5 = through(not(clborder), bSel5)
let b6 = through(not(clborder), bSel6)
let b7 = through(not(clborder), bSel7)
let b8 = through(not(clborder), bSel8)
let b9 = through(not(clborder), bSel9)
let b10 = through(not(clborder), bSel10)
let b11 = through(not(clborder), bSel11)
    
// save branches' selected vertices -----------------------------------------------------
save "root (selected verts)" rootSel
save "b0 (selected verts)" bSel0
save "b1 (selected verts)" bSel1
save "b2 (selected verts)" bSel2
save "b3 (selected verts)" bSel3
save "b4 (selected verts)" bSel4
save "b5 (selected verts)" bSel5
save "b6 (selected verts)" bSel6
save "b7 (selected verts)" bSel7
save "b8 (selected verts)" bSel8
save "b9 (selected verts)" bSel9
save "b10 (selected verts)" bSel10
save "b11 (selected verts)" bSel11

// save computed branches ---------------------------------------------------------------
save "root" root
save "b0" b0
save "b1" b1
save "b2" b2
save "b3" b3
save "b4" b4
save "b5" b5
save "b6" b6
save "b7" b7
save "b8" b8
save "b9" b9
save "b10" b10
save "b11" b11

// save border property -----------------------------------------------------------------
save "clborder" clborder
\end{verbatim}
}

Specification in \imgql\ to find the branching hierarchy in a tree coral.

{\small
\begin{verbatim}
// all the cells which have a certain property x and can reach cells with property y
// through a path of cells that satisfy property y
let reach(x,y) = or(x, through(x,y))

// define near as classical closure derived from through
let near(x) = through(x,tt)

// all the simplexes which have a certain property x or have a property y and can
// reach simplexes with property x only through simplexes of property y
let grow(x,y) = or(x, and(y, through(y, x)))

// all simplexes which have either property x or property y but not both
//let xor(x,y) = or( and(x, not(y)), and(not(x), y) )

// set operators
let complement(x) = not(x)
let union(x,y) = or(x,y)
let intersection(x,y) = and(x,y)
let difference(x,y) = and(x, not(y))
let symmetricDifference(x,y) = xor(x,y)

//code start
load model = "./polylogica/CORAL_Mke/EDR26_T6_with_branches.json"


// face properties ----------------------------------------------------------------------
// we expand them to the adjacent vertices and edges using the closure operator
let rank1 = near(ap("rank1"))
let rank2 = near(ap("rank2"))
let rank3 = near(ap("rank3"))
let border = near(ap("border"))

// vertex properties --------------------------------------------------------------------
// here a subset of vertices for the root branch has been manually marked and it is used
// to identify the set of simplexes belonging to the root branch.
let root = through(not(border), ap("root"))
    
// compute rank order through PolyLogicA ----------------------------------------------
// we define two properties for each rank level
// the first property describes all the cells which belong to the given rank level
// the second property describes all the cells which belong to the given rank level
// or lower, also containing the borders between them.
let rankP1 = root
let rankA1 = root

// rank level n+1 defines as all the simplexes for which there exists a path that crosses
// only one border and then reaches any cell belonging to rank level n. Additionally
// the cells in rank level n+1 cannot be borders or belong to any other rank level.
// the border between two rank levels consists of those border cells which can reach
// both rank levels.
let through2(first, second, end) = through(first, grow(end, second))
let rankP2 = near(difference(through2(not(border), border, rankP1), rankA1))
let rankA2 = or(rankP2, grow(rankA1, border))
let rankP3 = near(difference(through2(not(border), border, rankP2), rankA2))
let rankA3 = or(rankP3, grow(rankA2, border))

// identify which branches do not have children -----------------------------------------
// the branches which do not have children are those for which there is not path that
// reaches an higher rank order passing through only one border. 
let leafsP = or(difference(rankP1, grow(grow(rankP2, border), rankP1)), 
             or(difference(rankP2, grow(grow(rankP3, border), rankP2)), rankP3))

// save border and rank properties ------------------------------------------------------
save "border" border
save "rank1 (PolyLogicA)" rankP1
save "rank2 (PolyLogicA)" rankP2
save "rank3 (PolyLogicA)" rankP3
\end{verbatim}
}

\subsection{Full \imgql\ Specification for the Aircraft Example}\label{app:ufo}

{\small
\begin{verbatim}
// standard library ---------------------------------------------------------------------

// Direct Near and Converse Near
let dirnear(x) = through(x,tt)
let cnvnear(x) = cvnear(x) 

// all the cells which have a certain property x or than can reach property y
// through a path of cells that satisfy property x
let reach(x,y) = and(x, through(x,y))

// all the cells which have a certain property x or have a property y and can
// reach cells with property x only through cells with property y
let grow(x,y) = or(x, and(y, through(y, x)))

// surround
let sur(x,y) = and(x, not(through(not(y),not(or(x,y)))))

// all cells which have either property x or property y but not both
//let xor(x,y) = or( and(x, not(y)), and(not(x), y) )

// set operators
let complement(x) = not(x)
let union(x,y) = or(x,y)
let intersection(x,y) = and(x,y)
let difference(x,y) = and(x, not(y))
let symmetricDifference(x,y) = xor(x,y)

//code start
load model = "polylogica/UFO_curvature_final/13885_UFO_Triangle_V5_L2_MKE_curvature_more.json"

//UFO properties
let flat = (ap("flat"))
let convex = (ap("convex"))
let convexsharp = (ap("convexsharp"))
let concave = ap("concave")
let ufo = ap("ufo")

let moreflat = cnvnear(flat)
let moreconvex = cnvnear(convex)
let moreconvexsharp = cnvnear(convexsharp)
let moreconcave = cnvnear(concave)

// Convex vertices that are close to concave ones
let convexmeetsconcave = convex & through(moreconcave,convex)
// Areas with a small ridge: concave vertices and sharp convex ones are near to each other
let ridgeborder = concave & through(moreconvexsharp,concave)
// extend the ridge border with adjacent edges and triangles
let moreridgeborder = cnvnear(ridgeborder)
// not flat area surrounded by a ridge: very thin areas along the wings and front of tail 
let ridgearea = ufo & sur((moreconvex | moreconvexsharp | ufo),moreridgeborder)
// flat area surrounded by a ridge (triangle at bottom and areas at back of tail)
let flatridgearea = moreflat & (!grow((ridgeborder & concave), moreflat))


// Results:
save "flat" flat
save "convex" convex
save "convexsharp" convexsharp
save "concave" concave
save "ufo" ufo
save "moreflat" moreflat
save "moreconvex" moreconvex
save "moreconvexsharp" moreconvexsharp
save "moreconcave" moreconcave
save "convexmeetsconcave" convexmeetsconcave
save "ridgeborder" ridgeborder
save "moreridgeborder" moreridgeborder
save "ridgearea" ridgearea
save "flatridgearea" flatridgearea
\end{verbatim}
}

\subsection{Script to Convert Images in the Wavefront .obj Format into the Input Format for \polylogica{}}
\label{sec:convert}

Example python script for the coral example.

{\small
\begin{verbatim}
## obj to imgql file
import sys
import os
import json

# modify this code to change how atomic propositions are loaded
object_scale = 50
atom_names = ["hack1", "hack2", "hack3", "border", "root"]
for i in range(13): atom_names.append('b' + str(i))

def is_equal_with_error(left, right, error):
    return abs(left - right) < error

def is_color(properties, color):
    return is_equal_with_error(properties['r'], color[0], 10) 
                and is_equal_with_error(properties['g'], color[1], 10)
                and is_equal_with_error(properties['b'], color[2], 10)

def get_vertex_atoms(vertex):
    if is_color(vertex, [255, 255, 255]):
        return ["root", "b0"]
    elif is_color(vertex, [100, 255, 255]):
        return ["b1"]
    elif is_color(vertex, [255, 100, 255]):
        return ["b2"]
    elif is_color(vertex, [255, 255, 100]):
        return ["b3"]
    elif is_color(vertex, [100, 100, 255]):
        return ["b4"]
    elif is_color(vertex, [100, 255, 100]):
        return ["b5"]
    elif is_color(vertex, [255, 100, 100]):
        return ["b6"]
    elif is_color(vertex, [50, 255, 255]):
        return ["b7"]
    elif is_color(vertex, [255, 50, 255]):
        return ["b8"]
    elif is_color(vertex, [255, 255, 50]):
        return ["b9"]
    elif is_color(vertex, [50, 50, 255]):
        return ["b10"]
    elif is_color(vertex, [50, 255, 50]):
        return ["b11"]
    elif is_color(vertex, [255, 50, 50]):
        return ["b12"]
    else:
        return []

def get_edge_atoms(edge):
    return []

def get_triangle_atoms(triangle):
    if is_color(triangle, [12, 7, 133]):
        return ["hack1"]
    elif is_color(triangle, [107, 0, 167]):
        return ["hack2"]
    elif is_color(triangle, [179, 45, 140]):
        return ["hack3"]
    elif is_color(triangle, [100, 100, 100]):
        return ["border"]
    else:
        return []

# the actual code starts here
# first parse the materials file, if exists
mtl_data = []
with open(sys.argv[1] + '.mtl') as mtl_file:
    mtl_data = mtl_file.readlines()
mtl_data = [ line.strip() for line in mtl_data ]

def float_color_to_8bit(color):
    return int(round(float(color) * 255))

materials = {}
current_material = { 'name': None }
for line in mtl_data:
    line = line.split()
    if len(line) == 0:
        continue
    if line[0] == "newmtl":
        if current_material['name'] != None:
            materials[ current_material['name'] ] = current_material
            current_material = {}
        current_material['name'] = line[1]
    elif line[0] == "Ka":
        current_material['ambient'] = [ float_color_to_8bit(line[i + 1]) for i in range(3) ]
    elif line[0] == "Kd":
        current_material['diffuse'] = [ float_color_to_8bit(line[i + 1]) for i in range(3) ]
    elif line[0] == "Ks":
        current_material['specular'] = [ float_color_to_8bit(line[i + 1]) for i in range(3) ]

#push the last material
if current_material['name'] != None:
    materials[ current_material['name'] ] = current_material

# the obj file can now be parsed
obj_data = []
with open(sys.argv[1]) as obj_file:
    obj_data = obj_file.readlines()
obj_data = [ line.strip() for line in obj_data ]

# collect the points
points = []
idCounter = 0
for line in obj_data:
    line = line.split()
    if len(line) == 0:
        continue
    if line[0] == "v":
        point = {}
        point['x'] = float(line[1]) * object_scale
        point['y'] = float(line[2]) * object_scale
        point['z'] = float(line[3]) * object_scale
        if len(line) == 7:
            point['r'] = float_color_to_8bit(line[4])
            point['g'] = float_color_to_8bit(line[5])
            point['b'] = float_color_to_8bit(line[6])
        points.append(point)

# collect the edges and triangles
edges = {}
triangles = []
current_material = { "r": 0, "g": 0, "b": 0 }
for line in obj_data:
    line = line.split()
    if len(line) == 0:
        continue
    if line[0] == "f":
        vertices = {}
        for i in range(3):
            vertex_index = int(line[ i + 1 ].split('//')[0]) - 1
            vertices['V' + str(i)] = vertex_index

        for (i,j) in [(0,1),(1,2),(2,0)]:
            edge = {}
            edge['V0'] = vertices['V' + str(i)]
            edge['V1'] = vertices['V' + str(j)]
            if edge['V0'] > edge['V1']:
                edge['V0'], edge['V1'] = edge['V1'], edge['V0']
            name = str(edge['V0']) + '-' + str(edge['V1'])
            if name not in edges:
                edges[name] = edge

        triangle_data = vertices | current_material
        triangles.append(triangle_data)
    elif line[0] == "usemtl":
        current_material_name = line[1]
        current_material = {
            "r": materials[current_material_name]['diffuse'][0],
            "g": materials[current_material_name]['diffuse'][1],
            "b": materials[current_material_name]['diffuse'][2]
        }

edges = [ v for k, v in edges.items() ]

def simplexItem(id, points, atoms):
    result = {}
    result['id'] = id
    result['points'] = points
    result['atoms'] = atoms
    return result

# create and fill json file
json_data = {}
json_data['atomNames'] = atom_names
json_data['numberOfPoints'] = len(points)
json_data['coordinatesOfPoints'] = [ [point['x'], point['y'], point['z']] for point in points ]

points_data = []
for id, point in enumerate(points):
    point_item = simplexItem(
        'P' + str(id),
        [ id ],
        get_vertex_atoms(point)
    )
    points_data.append(point_item)

edge_data = []
for id, edge in enumerate(edges):
    edge_item = simplexItem(
        'E' + str(id),
        [ edge['V0'], edge['V1'] ],
        get_edge_atoms(edge)
    )
    edge_data.append(edge_item)

face_data = []
for id, face in enumerate(triangles):
    face_item = simplexItem(
        'T' + str(id),
        [ face['V0'], face['V1'], face['V2'] ],
        get_triangle_atoms(face)
    )
    face_data.append(face_item)

json_data['simplexes'] = points_data + edge_data + face_data

out_file_name = os.path.splitext(sys.argv[1])[0] + '.json'
json.dump(json_data, open(out_file_name, "w"), indent=2)

\end{verbatim}
}

\end{document}